\RequirePackage{fix-cm}
\documentclass[twocolumn,10p]{svjour3}          % twocolumn
\smartqed  % flush right qed marks, e.g. at end of proof
\usepackage{amssymb}
\usepackage{amsmath}
\usepackage{wasysym}
\usepackage{natbib}
\usepackage{caption}
\usepackage{graphicx}
\usepackage{epstopdf}
\usepackage{url}
\usepackage{marvosym}
\usepackage{epsfig}
\usepackage{subcaption}
\usepackage{tikz}
\usetikzlibrary{arrows,calc,intersections,positioning,arrows,decorations.pathmorphing,decorations.markings,shapes}%

\newcommand{\shorteq}{%
\settowidth{\@tempdima}{--}% Width of hyphen
\resizebox{\@tempdima}{\height}{=}%
}

\begin{document}

\title{Laser and Radio Tracking for Planetary Science Missions - A Comparison%\thanks{Grants or other notes
%about the article that should go on the front page should be
%placed here. General acknowledgments should be placed at the end of the article.}
}

%\titlerunning{Short form of title}        % if too long for running head

\author{Dominic Dirkx \and Ivan Prochazka \and Sven Bauer \and Pieter Visser \and Ron Noomen \and Leonid I. Gurvits \and Bert Vermeersen
}

%\authorrunning{Short form of author list} % if too long for running head

\institute{D. Dirkx \and P. Visser \and R. Noomen \and L.I. Gurvits \and B. Vermeersen \at
             Kluyverweg 1, 2629HS, Delft, the Netherlands \\
              Tel.: +31(0)15-2788866\\
              \email{d.dirkx@tudelft.nl}           %  \\
%             \emph{Present address:} of F. Author  %  if needed
           \and
           I. Prochazka \at
              Czech Technical University in Prague,  Brehova 7, 115 19 Prague 1, Czech Republic
          \and
          S. Bauer \at
        GeoForschungsZentrum Potsdam (GFZ), Dep. 1 Geodesy and Remote Sensing, Telegrafenberg, Potsdam, Germany
        \and
        L.I. Gurvits \at
       Joint Institute for VLBI ERIC,  PO Box 2, 7990 AA Dwingeloo, The Netherlands
}

\date{Received: date / Accepted: date}
% The correct dates will be entered by the editor

\maketitle

\begin{abstract}
At present, tracking data for planetary missions largely consists of radio observables: range-rate (Doppler), range and angular position (VLBI/$\Delta$DOR). Future planetary missions may use Interplanetary Laser Ranging (ILR) as a tracking observable. Two-way ILR will provide range data that are about 2 orders of magnitude more accurate than radio-based range data. ILR does not produce Doppler data, however. In this article, we compare the relative strength of radio Doppler and laser range data for the retrieval of parameters of interest in planetary missions, to clarify and quantify the science case of ILR, with a focus on geodetic observables.

We first provide an overview of the near-term attainable quality of ILR, in terms of both the realization of the observable and the models used to process the measurements. Subsequently, we analyze the sensitivity of radio-Doppler and laser-range measurements in representative mission scenarios for parameters of interest. We use both an analytical approximation and numerical analyses of the relative sensitivity of ILR and radio Doppler observables for more general cases.

We show that mm-precise range normal points are feasible for ILR, but mm-level accuracy and stability in the full analysis chain is unlikely to be attained, due to a combination of instrumental and model errors. We find that ILR has the potential for superior performance in observing signatures in the data with a characteristic period of greater than 0.33-1.65 hours (assuming 2-10 mm uncertainty for range and 10 $\mu$m/s at 60 s for Doppler). This indicates that Doppler tracking will typically remain the method of choice for gravity field determination and spacecraft orbit determination in planetary missions. ILR data will be able to supplement the orbiter tracking data used for the estimation of parameters with a once-per-orbit signal. Laser ranging data, however, are shown to have a significant advantage for the retrieval of rotational and tidal characteristics from landers. Similarly, laser ranging data will be superior for the construction of planetary ephemerides and the improvement of solar system tests of gravitation, both for orbiter and lander missions. 
 
% \PACS{PACS code1 \and PACS code2 \and more}
% \subclass{MSC code1 \and MSC code2 \and more}
\end{abstract}

\section{Introduction}
\label{sec:planetaryMissions}

For both Earth-orbiting and planetary missions, a variety of tracking observables is available from which the trajectory of the spacecraft can be reconstructed. % \citep[\textit{e.g.},][]{MontenbruckGill2000,Moyer2005,FiengaEtAl2009}. %The reconstruction of the trajectory of a spacecraft is crucial for a variety of reasons. 
A tracking data type that will become available for future planetary missions is Interplanetary Laser Ranging (ILR) %which is based on the transmission and reception of short (10-100 ps) laser pulses 
\citep{Degnan2002}. The technology for such a system derives strongly from Satellite Laser Ranging \citep[SLR; ][]{PearlmanEtAl2002}, Lunar Laser Ranging \citep[LLR; ][]{Murphy2013} as well as Laser Time Transfer \citep[LTT; ][]{ProchazkaEtAl2011,ExertierEtAl2014}. The key difference between SLR/LLR and ILR is that ILR requires a transponder on both the ground and space segments, as opposed to the purely passive space segment in SLR/LLR (retroreflectors).

Currently, deep-space missions largely rely on the use of radio tracking for their orbit determination and the associated parameter estimation. In particular Dopp\-ler data %which encode the velocity of the spacecraft projected along the line-of-sight to the ground station, 
has been the primary data type for this application,
%as well as for the estimation of associated quantities such as gravity field coefficients, body rotational properties, and tidal Love numbers
 \citep[\emph{e.g.} ][]{KonoplivEtAl2011,IessEtAl2012,MazaricoEtAl2014b}. Radio-based range data supplement the Doppler data by providing the absolute distance between spacecraft and ground station. %by using a set of radio telescopes 
The angular position of the spacecraft in the sky can be measured by means of Very Long Baseline Interferometry (VLBI) \citep{DuevEtAl2012}. {In the so-called PRIDE (Planetary Radio Interferometry and Doppler Experiments) setup, VLBI and Doppler data are obtained concurrently \citep{DuevEtAl2016}.} Unlike the Doppler data, range and VLBI data are used primarily for the estimation of solar system ephemerides \citep{FiengaEtAl2009, JonesEtAl2015, DirkxEtAl2017}, which provide crucial input for experimental relativity \citep[\emph{e.g.} ][]{Will2014}. The selection of ILR as a tracking type in future missions is contingent upon its data being able to provide scientific results that are complementary/supplementary to, or competitive with, the results obtained from existing systems, such as those mentioned above. %Analyzing the science cse of ILR for planetary science objectives is the primary goal of this article.

There has been a number of experimental demonstrations of ILR, both in one-way \citep{AbshireEtAl2006,NeumannEtAl2008,NodaEtAl2017} and two-way \citep{SmithEtAl2006} modes. The only operational implementation of ILR to date has been on the Lunar Reconnaissance Orbiter (LRO), {using the laser altimeter system} %was used to obtain one-way laser ranging data over a period of five years 
\citep{ZuberEtAl2010, BauerEtAl2016b, MaoEtAl2016}. None of the demonstrations of ILR have been performed with dedicated hardware, and the attainable measurement accuracy has not yet been pushed to the limit of state-of-the-art.
In recent years, there has been substantial development in LTT \citep{ExertierEtAl2014,SamainEtAl2015,ExertierEtAl2017,ProchazkaEtAl2017}, %, in particular for the Time Transfer by Laser Link (T2L2) and European Laser Timing (ELT) experiments, 
which is in many ways similar to a transponder ranging system.  %Additionally, (interplanetary) laser communications systems have a strong synergy with ILR \citep{HemmatiEtAl2009}, as demonstrated with the LLCD payload on the LADEE lunar orbiter. 
Laser communication technology is also maturing for use in planetary missions, as demonstrated by the LLCD demonstrator on the LADEE lunar orbiter. As an \emph{ad hoc} product, LLCD communications data were used to obtain two-way ranging data with a precision of several cm \citep{StevensEtAl2016}. 

%as well as interplanetary laser communications \citep{StevensEtAl2016}. These technologies have strong synergy with ILR, from both a hardware and operations point-of-view.

%LTT experiments in Earth orbit are ongoing. The Time Transfer by Laser Link (T2L2) experiment onboard the Jason-2 satellite continues to provide high-quality data \citep{ExertierEtAl2014}. Developments for T2L2, as well as the upcoming Europeal Laser Timing (ELT) experiment \citep{ProchazkaEtAl2012}, are invaluable for the future development of ILR \citep{SamainEtAl2015,ExertierEtAl2017,ProchazkaEtAl2017}, both from a hardware development and from an operations point-of-view. Additionally, (interplanetary) laser communications systems have a strong synergy with ILR \citep{HemmatiEtAl2009}, as demonstrated with the LLCD payload on the LADEE spacecraft, which performed laser communications from lunar orbit. As an \emph{ad hoc} product, LLCD communications data were used to obtain two-way ranging data with a precision of several cm \citep{StevensEtAl2016}. 

Thus far, analyses of various aspects of ILR have focussed on single mission concepts \citep{TuryshevEtAl2004,ChandlerEtAl2005,MerkowitzEtAl2007,LuoEtAl2009,ChristopheEtAl2009,TuryshevEtAl2010, BirnbaumEtAl2010,OberstEtAl2012,Iorio2013b,DirkxEtAl2014}, or operational and data analysis aspects % associated with its implementation
\citep{FolknerFinger1990,Degnan1996,Degnan2002,Degnan2008,SchreiberEtAl2009,DirkxEtAl2014b,DirkxEtAl2015, DirkxEtAl2015b, BauerEtAl2016b,BauerEtAl2016a, MaoEtAl2016}. Additionally, the majority of these analyses has focused on using the laser ranging data to improve solar system tests of relativity. Considering the great wealth of information on \emph{e.g.} planetary geodesy that radio tracking has provided, it is natural to extend the analysis of ILR to planetary science objectives.%, as has been done for a Phobos lander by \cite{TuryshevEtAl2010, DirkxEtAl2014b}. 

Our goal is to provide a comparison of radio-Doppler and laser-range data for planetary missions, {and} to identify the areas where the addition of ILR data would be beneficial for attaining the scientific objectives of planetary missions. A preliminary set of results is given by \cite{Dirkx2015}. Instead of focusing on a single mission concept in detail, we take a broad view and quantitatively analyze the relative strength of the range and Doppler observables for parameters of interest in a variety of planetary missions. As a result, we identify classes of science products/mission scenarios for which ILR will be a competitive design option. %, and provide an estimate of the improvement that ILR data could provide compared to radio tracking of a comparable mission. Our focus is on the comparison to Doppler data.

We {give an overview of tracking data in Section \ref{sec:planetaryTrackingData} and discuss} the expected error budget of ILR in Section \ref{sec:lrData}. In Section \ref{sec:compCrit}, we present the methods we use to compare ILR and Doppler data. We use both an analytical {approach}, and a numerical covariance analysis based on simulated orbit determination/parameter estimation. The numerical technique serves to identify the level of applicability of the analytical approach, and to provide guidelines on how to apply it in mission design and analysis. In Section \ref{sec:results} we show the results of our comparison of the two data types. In Section \ref{sec:discussion}, we discuss the implications {of these results for the use of ILR data} %the added value of ILR data to the main classes of science products that are obtained from planetary tracking data, 
, with a focus on geodetic observables. We conclude with a discussion on the overall science case of ILR in Section \ref{sec:conclusions}.

\section{Planetary Tracking Data}
\label{sec:planetaryTrackingData}

In this section, we give a general overview of planetary tracking data, including the models for both range and Doppler observables. %We first discuss the error budgets of typical radio tracking techniques in Section \ref{sec:radiometricTrackingDataQuality}. Subsequently, we discuss the general operating concept of ILR in Section \ref{sec:ilrConcepts}, followed by the mathematical description of the observables in Section \ref{sec:trackingTypes}. We conclude with a preliminary comparison between range and range-rate observables in Section \ref{sec:preliminaryTypeComparison}.

\subsection{Radio Data}
\label{sec:radiometricTrackingDataQuality}

%Here, we give a brief overview of the current quality of two-way radiometric tracking data for planetary missions.
%The primary observable for interplanetary orbit determination is the Doppler measurement, which encodes the time-averaged range-rate of the space segment along the line-of-sight to the ground station. Measurements of the absolute range between Earth and the target are also routinely obtained, which are primarily used as input to solar system ephemerides \citep{FiengaEtAl2009}. %which are realized by measuring the light time of a coded radio signal to obtain the direct range  between a ground station and the space segment \citep{}.
%In addition, an angular position measurement, realized by Very Long Baseline Interferometry (VLBI) techniques, is used as a supporting data type \citep{LanyiEtAl2007,DuevEtAl2012} to determine the out-of-plane component of ephemerides, as well as to fix the ITRF to the ICRF.  %Radio tracking of modern space missions is typically done in X-band ($\sim$7-8.5 GHz) and/or Ka-band ($\sim$32-35 GHz) \citep{ThorntonBorder2000}.

%In VLBI, a set of radio telescopes on Earth is simultaneously used to observe the same radio signal originating from the spacecraft. After resolving the cycle ambiguity, \textit{i.e.}, determining the total signal propagation time difference for the different stations, an angular observable may be obtained \citep{LanyiEtAl2007,DuevEtAl2012}. 

The typical precisions of radio tracking data %, as quantified by the postfit estimation residuals, 
are currently at the level of 0.02-0.1 mm/s at 60 s integration time for range-rate measurements at X-band \citep[\textit{e.g.},][]{ThorntonBorder2000,MartyEtAl2009,KonoplivEtAl2011,IessEtAl2014,BocanegraEtAl2017} and 0.5-5 m for range measurements \citep[\textit{e.g.},][]{ThorntonBorder2000,FolknerEtAl2014}.  Detailed discussions on sources of both systematic errors and random noise are given by \cite{ThorntonBorder2000,Moyer2005,AsmarEtAl2005,IessEtAl2014,CalvesEtAl2014}. 
%Both radiometric and interplanetary laser tracking can be performed in a one- or two-way configuration. For a one-way system, only the up- (or down-) link of the signal is used to obtain the measurement. However, this setup suffers from the fact that it typically introduces a large error in the processed observable, as it requires the comparison of the clocks onboard the ground and the space segment \citep{Moyer2005,AsmarEtAl2005}. For a two-way observable this effect is largely mitigated and typically does not have a dominant role in the error budget of radio-tracking data. %A two-way measurement uses both the up- and downlink of the signal to generate a measurement, so that the space- and ground-based clocks no longer need to be compared in an absolute manner.
For Doppler measurements, the systematic errors are typically close to negligible \citep[\emph{e.g.}][]{IessEtAl2014}. %, although small low-frequency systematic errors (periods on the order of 1,000-100,000 s) may be present. %Despite the typical absence of short-periodic error source in Doppler tracking, an anomalous signal in the Doppler measurements (at 1 s integration time) of MRO is found by \cite{GenovaEtAl2015}, who obtain an improvement in postfit residual of a factor of 2-3 after filtering this anomalous behaviour. Whether the source of this error is due to a modelling error or an inherent noise source in the data remains to be investigated. 
In contrast, the level of systematic errors of radio range measurements can be quite large, comparable to the random noise, at the 1 m level. %For situations with a small solar separation angle, this value increases to tens of meters and above, as the influence of the interplanetary plasma becomes dominant. 

{The radio tracking noise is dominated by the propagation effects in the interplanetary medium, and depend strongly on the solar elongation angle. For Mars Express, \cite{DuevEtAl2016} show that PRIDE Doppler noise at X-band is indeed dominated by the these effects, with a median one-way value of $\sim 30\mu$/s at 10 s integration time.}
Combining observations at multiple wave\-lengths allows the removal of the majority of these errors \citep{ReasenbergEtAl1979, BertottiEtAl2003}.
%This calibration was performed with ranging measurements at S- and X- band by the Viking landers \citep{ReasenbergEtAl1979} and Voyager spacecraft \citep{KrisherEtAl1991}. \cite{BertottiEtAl1993} proposed the use of multi-frequency Doppler measurements, as opposed to previous experiments with dual-frequency range data. 
%This approach was demonstrated with a Ka/X-band link using the instrumentation onboard the Cassini spacecraft \citep{BertottiEtAl2003} obtaining a Doppler noise level close to 1 $\mu$m/s level in some cases  at small solar separation angles (when scaled to 1000 s integration time).
%\footnote{Unfortunately, the Ka-band transponder on Cassini suffered a malfunction, preventing plasma noise cancellation on the downlink during the Saturn phase.}. %providing a much improved estimate of the relativistic parameter $\gamma$ (see Section \ref{sec:experimentalRelativity}).
%The use of multiple wavelengths largely mitigates the issue of tracking at small solar separation angles, where the data quality of single-frequency systems degrades due to the larger influence of the Sun's plasma environment. 
The combined X- and Ka-band approach {was/is} used for Cassini  {\citep{KlioreEtAl2004} and Juno \citep{IessEtAl2018}}, %\footnote{Unfortunately, the Ka-band transponder on Cassini suffered a malfunction, preventing plasma noise cancellation on the downlink during the Saturn phase.} 
and  {is anticipated for use} on the BepiColombo and %\citep{IessEtAl2009}, 
JUICE
% \citep{TommeiEtAl2015} 
missions, which have tracking data quality requirements of 0.01 mm/s at 60 s integration time and 3 $\mu$m/s at 1000 s integration time. %However, as discussed bove experience with the dual-freuqency link on Cassini has shown data noise levels down to 1 $\mu$m/s at 1000 s, making it likely that these future missions will be able to go beyond their required 3 $\mu$m/s range-rate requirement. 
Additionally, these missions will use an advanced radio ranging system, allowing two-way range measurements with a predicted accuracy of down to 20 cm.

\subsection{Laser Ranging - Measurement Concepts}
\label{sec:ilrConcepts}
SLR has been used for Earth-orbiting satellites for more than 50 years \citep{PearlmanEtAl2002}, and provides two-way range data with sub-cm accuracy. %In SLR, short (10-100 ps) laser pulses are transmitted by ground stations to these spacecraft. The pulses are reflected back to the ground station, where the pulse time-of-flight is measured, allowing a two-way range measurement to be obtained.
% with sub-cm accuracy. %The technique provides range measurements with accuracy at the cm-level and is one of four space-geodetic techniques used as input to the realization of the ITRF \citep{AltamimiEtAl2011}. Additionally, SLR is used for the determination of low-degree gravity field coefficients of the Earth \citep{Sosnica2014} and as a means of validation for spacecraft orbits determined using GNSS \citep{UrschlEtAl2005}.
In ILR, the use of retroreflectors is no longer feasible due to the large target distance %as the received signal intensity is proportional to the inverse fourth power of the target distance 
%\citep{Degnan1995}. To overcome this limitation,
{requiring an active system} in both the ground and space segments.%, the use of which changes the signal strength dependency to inverse square with distance. 

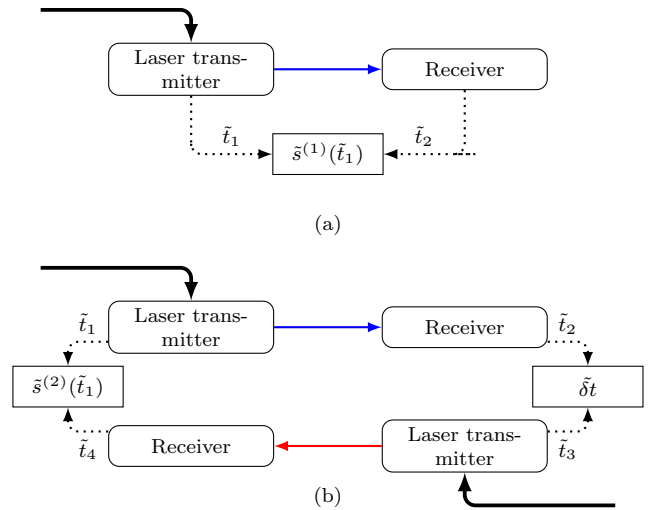
\begin{figure}[tb!]
\centering
\pgfdeclarelayer{background}
\pgfdeclarelayer{backbackground}
\pgfsetlayers{backbackground,background,main}
\footnotesize
% Define a few styles and constants
\tikzstyle{function}=[draw, fill=white!20, text width=2.2cm, text centered, rounded corners, minimum height=0.6cm, minimum width = 1.5cm ]
\tikzstyle{function2}=[draw, fill=white!20, text width=1.4cm, text centered, minimum height=0.6cm, minimum width = 1.2cm ]

\def\blockdist{4cm}
\def\funcdist{2.5cm}
\def\vertblockdist{1.75cm}

\begin{tikzpicture}[scale=0.9, every node/.style={transform shape}]

\begin{scope}[shift={(0,0)}]

\node [function](transmitterA){Laser transmitter};
\node [function, right of = transmitterA, node distance = \blockdist ](receiverA){Receiver};
\draw [-latex,rounded corners,color=blue,thick] (transmitterA) -- (receiverA);

\node [coordinate, above of = transmitterA, node distance = 0.5*\vertblockdist](a){};
\node [coordinate, left of = a, node distance =  0.1*\blockdist](b){};
\node [coordinate, left of = a, node distance =  0.55*\blockdist](c){};
\draw [-latex,line width=0.5mm,rounded corners](c) -- (b)-|(transmitterA.90);

\node [coordinate, below of = transmitterA, node distance = 0.5*\funcdist](intermediateA){};
\node [coordinate, below of = receiverA, node distance = 0.5*\funcdist](intermediateB){};

\node [function2, right of = intermediateA, node distance = 0.5*\blockdist](rangeBlock){$\tilde{s}^{(1)}(\tilde{t}_{1})$};

\draw [-latex,rounded corners,dotted,thick](transmitterA.270) |- (intermediateA) -- (rangeBlock)node[midway,above ] () {$\tilde{t}_{1}$};
\draw [-latex,rounded corners,dotted,thick] (receiverA.270)|- (intermediateB) -- (rangeBlock) node[midway,above ] () {$\tilde{t}_{2}$};

\begin{scope}[shift={(0.5*\blockdist,-2.3cm)}]
\node [draw=none](0,0){(a)};
\end{scope}

\end{scope}

\begin{scope}[shift={(0,-3.8cm)}]

\node [function](transmitterB){Laser transmitter};
\node [function, right of = transmitterB, node distance = \blockdist ](receiverB){Receiver};
\draw [-latex,rounded corners,color=blue,thick] (transmitterB) -- (receiverB);

\node [coordinate, above of = transmitterB, node distance = 0.5*\vertblockdist](a){};
\node [coordinate, left of = a, node distance =  0.1*\blockdist](b){};
\node [coordinate, left of = a, node distance =  0.55*\blockdist](c){};
\draw [-latex,line width=0.5mm,rounded corners](c) -- (b)-|(transmitterB.90);

\node [function, below of = receiverB, node distance = \vertblockdist](transmitterB2){Laser transmitter};

\node [coordinate, below of = transmitterB2, node distance = 0.5*\vertblockdist](a){};
\node [coordinate, right of = a, node distance =  0.1*\blockdist](b){};
\node [coordinate, right of = a, node distance =  0.55*\blockdist](c){};
\draw [-latex,line width=0.5mm,rounded corners](c) -- (b)-|(transmitterB2.270);

\node [function, left of = transmitterB2, node distance = \blockdist ](receiverB2){Receiver};
\draw [-latex,rounded corners,color=red,thick] (transmitterB2) -- (receiverB2);

\node [coordinate, left of = transmitterB, node distance = 0.45*\blockdist](intermediateA){};
\node [coordinate, left of = receiverB2, node distance = 0.45*\blockdist](intermediateB){};

\node [function2, below of = intermediateA, node distance = 0.5*\vertblockdist](rangeBlock){$\tilde{s}^{(2)}(\tilde{t}_{1})$};

\draw [-latex,rounded corners,dotted,thick](transmitterB.190) -| (rangeBlock)node[near start,above] () {$\tilde{t}_{1}$};
\draw [-latex,rounded corners,dotted,thick](receiverB2.170) -| (rangeBlock)node[near start,below] () {$\tilde{t}_{4}$};

\node [coordinate, right of = receiverB, node distance = 0.45*\blockdist](intermediateC){};
\node [coordinate, right of = transmitterB2, node distance = 0.45*\blockdist](intermediateD){};

\node [function2, below of = intermediateC, node distance = 0.5*\vertblockdist](dtBlock){$\tilde{\delta t}$};

\draw [-latex,rounded corners,dotted,thick](receiverB.350) -| (dtBlock)node[near start,above] () {$\tilde{t}_{2}$};
\draw [-latex,rounded corners,dotted,thick](transmitterB2.10) -| (dtBlock)node[near start,below] () {$\tilde{t}_{3}$};

\begin{scope}[shift={(0.5*\blockdist,-2.5cm)}]
\node [draw=none](0,0){(b)};
\end{scope}

\end{scope}

\end{tikzpicture}
\caption{Schematic representation of laser ranging concepts. Thick lines indicate signals to trigger the transmission of a laser pulse. a) One-way laser ranging b) Two-way asynchronous laser ranging. Figure adapted from \cite{Dirkx2015}.}
\label{fig:laserRangingConcepts}
\end{figure}

Two main types of active laser ranging systems (sometimes termed transponder laser ranging) can be distinguished for use in planetary missions \citep[][see Fig. \ref{fig:laserRangingConcepts}]{Degnan2002,BirnbaumEtAl2010}:

\begin{itemize}
\item One-way laser ranging. A laser pulse is transmitted from a ground station and detected by a (satellite-based) receiver (or the other way around). %The pulse transmission time $\tilde{t}_{t}$ at the ground station and reception time at the space segment $\tilde{t}_{r}$ are then used to determine the one-way range from Eq. (\ref{eq:rawTwoWayRange}).
An important issue with this method is that the transmission and reception times are measured by different, unsynchronized, clocks \citep{Bauer2017}. %As a result, offsets and rate differences between transmitter and receiver clocks will introduce an error in the range determination. 

\item Two-way asynchronous laser ranging. Both the space and ground segment fire laser pulses towards one another independently. %This negates the problem that occurs in the echo laser ranging concept, where laser pulses are sent in response to a noise signal. 
By pairing a range measurement from the up- and down-link, a two-way range measurement is obtained \citep{BirnbaumEtAl2010}, which does not suffer from the clock error issue of a one-way system. 
\end{itemize}

%For the one-way system, a one-way range $s^{(1)}(t)$ is measured, in which $t_{t}=t_{1}$ and $t_{r}=t_{2}$ in Eq. (\ref{eq:processedTwoWayRangePre}). For the two-way systems, it is the two-way range $s^{(2)}(t)$ that is measured, with $t_{t}=t_{1}$ and $t_{r}=t_{4}$ in Eq. (\ref{eq:processedTwoWayRangePre}). For active two-way ranging, the difference between uplink reception time $t_{2}$ and downlink transmission time $t_{3}$, denoted by $\delta t$ and termed the retransmission delay in an echo system, is measured by the space segment as:
%\begin{align}
%\delta t=t_{3}-t_{2} 
%\end{align}
%For typical SLR/LLR operations, the $\delta t$ term is absorbed into the retroreflector signature, \citep[ \textit{e.g.},][]{OtsuboAppleby2003}. For the two-way transponder concepts, the times $t_{2}$ and $t_{3}$ are measured and $\delta t$ is included in the observation model directly. More details on the mathematical formulation of the one- and two-way range observations are given in Section 3.2 of Chapter \ref{sec:paper3}. 

Two-way asynchronous laser ranging has been the me\-thod of choice in most ILR mission proposals%\citep[\emph{e.g. }][]{MerkowitzEtAl2007,TuryshevEtAl2010,OberstEtAl2012}
, due to its higher data quality and less stringent clock requirements. In this article our main focus is on two-way systems. % as it has neither the clock synchronization problem of a one-way system, nor the sensitivity to high noise levels of the echo system. 
%However, a one-way system is easier to operate and potentially cheaper to build and fly, since it does not require a laser generation and transmission system. Exploiting one-way data does impose more stringent requirements on clock stability, however. 

%A notable example of the use of a one-way system is the LRO spacecraft, for which one-way laser ranging was employed to support precise orbit determination \citep{ZuberEtAl2010}. 

Both the one- and two-way observable are created from time tags of the transmission ($\tau_{t}$) and reception ($\tau_{r}$), in their local proper time scales\footnote{In the context of this article, we use the symbol $t$ to denote either a coordinate time such as Barycentric Coordinate Time (TCB) or a scaled coordinate time such as Dynamical Barycentric Time (TDB). Although TDB is decidedly not a coordinate time, as discussed in detail by \cite{Klioner2008}, the distinction is not relevant for the purpose of our discussion. Details of the conversion between coordinate and proper time scales are discussed by \cite{SoffelEtAl2003}, while the influence on the data analysis is discussed by \cite{DirkxEtAl2015b}.}.
The error-free two-way raw range measurement is then created from the coordinate times $t$  as:
\begin{align}
s(t_{t})=c\left(t_{r}-t_{t}\right)\label{eq:basicIdealRange}
\end{align}
%Eq. (\ref{eq:basicIdealRange}) represents the ideal range measurement. However, 
Due to uncalibrated errors in the measurements (Section \ref{sec:lrData}) of $\tau_{t}$ and $\tau_{r}$, the range measurement quality is degraded. We denote the measured time (transformed to coordinate time) as $\tilde{t}$, so that the measured range $\tilde{s}$ becomes:
\begin{align}
\tilde{s}(\tilde{t}_{t})&=c(\tilde{t}_{r}-\tilde{t}_{t})\label{eq:rawTwoWayRange}\\
&=s(t_{t})+\varepsilon_{s}\label{eq:rawTwoWayRange2}
\end{align}
where $\varepsilon_{s}$ is the lumped range error. Note that this range value includes the effects of atmospheric and relativistic effects.%From Eq. (\ref{eq:rawTwoWayRange}), it can be seen that constant errors in signal timing $t_{t}$ and $t_{r}$ will drop out off the two-way range measurement, making it insensitive to absolute timing errors. 

%Routine calibration at SLR ground stations, \citep[\textit{e.g.},][]{KirchnerEtAl2014}, allow part of the term $\epsilon_{s}$ to be resolved before data delivery. We denote this \textit{in situ} determination of the range bias as $\hat{\epsilon}_{s}$. This results in the following measurement that is used in the orbit determination, denoted $\hat{s}(t_{t})$:
%\begin{align}
%\hat{s}(\tilde{t}_{t})&=\tilde{s}(\tilde{t}_{t})-\hat{\epsilon}_{s}\label{eq:processedTwoWayRangePre}\\
%&={s}(t_{t})+(\epsilon_{s}-\hat{\epsilon}_{s})\label{eq:processedTwoWayRange}
%\end{align}
%so that the remaining range error is the difference between the true and \textit{a priori} estimated value of $\epsilon_{s}$.

\subsection{Observation Models}
\label{sec:trackingTypes}

We will denote the one-way range observation between point $A$ (transmitter) and point $B$ (receiver) by $s^{(1)}_{BA}$. {From their position functions, }denoted $\mathbf{r}_{A}(t)$ and $\mathbf{r}_{B}(t)$, respectively, the one-way range is obtained from:
\begin{align}
%s^{(1)}_{BA}(t_{t}{{=}} t_{1})=\left|\mathbf{r}_{B}\left(t_{1}+\frac{s^{(1)}_{BA}}{c}\right)-\mathbf{r}_{A}\left(t_{1}\right)\right|+\Delta s^{(1)}_{BA}(t_{1},t_{2})\label{eq:prelimRange}\\
s^{(1)}_{BA}(t_{r}{{=}} t_{2})=\left|\mathbf{r}_{B}\left(t_{2}\right)-\mathbf{r}_{A}\left(t_{2}-\frac{s^{(1)}_{BA}}{c}\right)\right|+\Delta s^{(1)}_{BA}(t_{1},t_{2})\label{eq:prelimRange2}
\end{align}
where the formulation is referenced to the
%transmission time $t_{t}$, here equal to $t_{1}$, and the second one to the 
reception time $t_{r}$, here equal to a given $t_{2}$. 
%That is, for the first equation, the transmission time $t_{1}$ is assumed to be known, and the reception time and resulting range measurement are calculated from this value. For the second equation, the reception time $t_{2}$ is assumed known. 
The term $\Delta s^{(1)}_{BA}$ denotes range corrections due to \textit{e.g.}, propagation medium and relativistic effects. %The two-way range observable is obtained from the combination of the up- and downlink of the signal, including a retransmission time/delay time \citep[\textit{e.g. }][]{Moyer2005, DirkxEtAl2015}.

%The Doppler observable is obtained by measuring the change in range to the spacecraft over a certain integration time.

For a one-way range-rate (or Doppler) observable, denoted here as $\dot{s}^{(1)}_{BA}$, with an integration time denoted by $\Delta t_{i}$, the observable is modelled as \citep{Moyer2005}:
\begin{align}
\dot{s}^{(1)}_{BA}(t_{r}{{=}} t_{4})&= \frac{s_{BA}^{(1)}(t_{r}{{=}} t_{4})-s_{BA}^{(1)}(t_{r}{{=}} t_{2} )}{\Delta t_{i}}\label{eq:prelimRangeRate}%\\
%t_{2}&=t_{4}-\Delta t_{i}\label{eq:integrationTimeRangeRate}\\
%\dot{s}^{(1)}_{BA}(t_{t}{{=}} t_{3}) &= \frac{s_{BA}^{(1)}(t_{t}{{=}} t_{3})-s_{BA}^{(1)}(t_{t}{{=}} t_{1} )}{\Delta t_{i}}\label{eq:prelimRangeRate2}\\
%t_{1}&=t_{3}-\Delta t_{i}\label{eq:integrationTimeRangeRate2}
\end{align}
%where we have reference the observation to the reception and transmission time (both of the second range measurement) in Eqs (\ref{eq:prelimRangeRate}) and (\ref{eq:prelimRangeRate2}), respectively. 
where $\Delta t_{i}=t_{4}-t_{2}$. Here, two one-way range observables with reception times $t_{4}$ and $t_{2}$ (and associated transmission times $t_{3}$ and $t_{1}$) are used. {Typical values of $\Delta t$ are 1-60 s, but may be $>1000$ s in certain cases.} %Typically, the range-rate measurement is integrated over a period $\Delta t_{i}=t_{4}-t_{2}$ of about 60 s. However, both smaller integration times such as 1-10 s  and larger integration times such as 1000 s may be used.%\footnote{We note that the model we use here is valid for so-called closed-loop Doppler observables, which is typically used in interplanetary spacecraft tracking. It is mathematically equivalent to the instantaneous (or open-loop) Doppler effect averaged over the integration time $\Delta t_{i}$ \citep{BocanegraEtAl2017}}. 

The two-way range is modelled as the combination of two one-way ranges:
\begin{align}
%s^{(2)}_{BA}(t_{t}{{=}} t_{1})&=s^{(1)}_{BA}(t_{t}{{=}} t_{1})+s^{(1)}_{AB}(t_{t}{{=}} t_{3})+c\delta t_{B}\label{paper3eq:twoWayModelA}\\
%t_{3}&=t_{1}+\frac{s^{(1)}_{BA}(t_{t}{{=}} t_{1})}{c}+\delta t_{B}\\
s^{(2)}_{BA}(t_{r}{{=}} t_{4})&=s^{(1)}_{AB}(t_{r}{{=}} t_{4})+s^{(1)}_{BA}(t_{r}{{=}} t_{2})+c\delta t_{B}\label{paper3eq:twoWayModelB}\\
t_{2}&=t_{4}-\frac{s^{(1)}_{AB}(t_{r}{{=}} t_{4})}{c}-\delta t_{B}
\end{align}
where $\delta t_{B}$ represents the delay between the reception and retransmission of the signal at station $B$,{ typically $<$ 1 ms for radio data \citep{BertoneEtAl2017}, up to the order of 1 minute for ILR \citep{DirkxEtAl2015}. }

\section{Laser ranging data - Error Sources}
\label{sec:lrData}
The error budgets of radio tracking systems are well understood and quantified (Section \ref{sec:radiometricTrackingDataQuality}). Here we analyze and discuss the various sources of error in ILR measurements/analysis. %Firstly, inherent stochasticity in the realization of the observable is discussed in Section \ref{sec:inherentStochastic}. Secondly, errors introduced by imperfections in the measurement systems are given in Section \ref{eq:measErrors}. Thirdly, even in the limit of perfect measurements, errors in the models used during the data analysis impact the quality of the science products, which we discuss in Section \ref{sec:modelErrorSources}. Finally, we summarize the overall expected quality of ILR, and key areas that required further development, in Section \ref{sec:ilrTotalError}.

\subsection{Inherent Measurement Uncertainty}
\label{sec:inherentStochastic}

{The primary measurement uncertainty of ILR is a convolution of two main contributors: the laser pulse profile and the detector impulse response.} For a pulse with a perfectly Gaussian temporal profile and single-photon intensity detection energies, the pulse profile introduces a purely Gaussian measurement error (Murphy, 2001; Dirkx et al., 2014a). The one-way range precision limit due to {only} this effect is 1.3 to 13 mm single-shot root mean square (RMS) for laser pulses with a full-width half maximum (FWHM) of 10 to 100 ps, respectively. {Note that the actual single-shot precision may be limited by the detector random error (Section \ref{eq:measErrors})}. 

{In cases where the number of detectable photons is larger than one, additional range biases at the several mm level may be incurred \citep{DirkxEtAl2014b}, if traditional threshold detection is used. \cite{Degnan2017} has proposed the use of centroid detection that could be used to overcome this bias, by combining all incoming detections from a single pulse into a single waveform. In ILR, however, detection energies are expected to be at the single-photon level.}

%Due to the finite length of a laser pulse, there is an inherent random uncertainty in the range measurement. %, since one can only statistically describe which photon(s) was(were) detected. 
%For a pulse with a perfectly Gaussian temporal profile and single-photon intensity detection energies, this introduces a purely Gaussian measurement error \citep{Murphy2001,DirkxEtAl2014b}. The one-way range precision limit due to {only} this effect is in the range of 1.3 to 13 mm single-shot root mean square (RMS) for laser pulses with a full-width half maximum (FWHM) of 10 to 100 ps (corresponding to pulse width RMS of 4.3-43 ps ), respectively. {Note that the actual single-shot precision is typically limited by the detector random error (Section \ref{eq:measErrors})}.

%In cases where $N_{p}>1$, with $N_{p}$ the number of detectable photons, additional range biases at the several mm level may be incurred \citep{DirkxEtAl2014b}. Typical ILR systems, however, will generally operate at the single-photon level due to the large target distance. %A multi-array detector can be used, which allows multiple single-photon detections from a single laser pulse \citep{TuryshevEtAl2010,Murphy2013}.%, as was also proposed for the PLR concept described by \cite{TuryshevEtAl2010}. 

%An additional inherent source of stochastic measurement error is atmospheric turbulence  \citep{Gardner1976}. 

The contribution due to the atmospheric turbulence \citep{Gardner1976} is in most cases below the 0.5 mm level \citep{KralEtAl2005} ({see Section \ref{sec:modelErrorSources} for discussion of troposphere model errors}).
%The much larger influence of errors in the non-turbulent troposphere models are discussed in Section \ref{sec:modelErrorSources}.}

%The precision of the combined measurement over $N$ pulses (normal points) is limited by time-correlated error sources, such as optical turbulence \citep{Gardner1976}, which is inherently stochastic and short-term instrumental instability (see Section \ref{eq:measErrors}). However, such uncertainties typically only become relevant well below the 1 mm level for state-of-the-art SLR systems, \citep{KralEtAl2005,BlazejEtAl2011}, allowing sub-mm normal point precision (but not accuracy) to be routinely achieved for single-photon ranging.

%As is the case for SLR, 
In SLR/LLR, the retroreflector signature causes significant distortion of the temporal pulse shape \citep{OtsuboAppleby2003}, which is not the case for ILR. %Since atmospheric effects do not significantly influence the laser pulse temporal energy distribution \citep{Fante1975, Degnan1995}, 
%As a result, temporal pulse shape at transmission and detection are approximately the same in ILR \citep{DirkxEtAl2014b}. 
Therefore, this aspect of the data stability could be better for ILR than for SLR, as {the incoming pulse energy temporal profile is more predictable in ILR.}% The uncertainty distribution of the pulse detection time is obtained from convolution of the pulse shape with detector impulse response (Section \ref{eq:measErrors})}.% which, for short laser pulses, is dominated by the detector, not the pulse size.}. %As a result, the single-shot inherent measurement uncertainty is directly defined by the size and shape of the laser pulse.

\subsection{Measurement Errors}
\label{eq:measErrors}
%Various sources of error in the measurement hardware degrade the quality of the measurements.
For ILR, the hardware-derived error sources are similar to those of SLR, which were summarized by \cite{ExertierEtAl2006}.  {Although much of the error budget remains close to that given there, we note several recent changes in the coming sections.} An important difference between SLR and ILR stems from the fact that in ILR part of the active hardware is on the space segment, and no passive reflectors are used.

%To assess the inaccuracies in both ground and space segment hardware for ILR, we rely in part on the extrapolation of existing ground stations. 
%A block diagram of a system is given in Fig. \ref{fig:detectionBlockDiagram}. 
For the characterization of the space segment, we rely in part on the development of space-grade detection systems that are currently in operation, such as T2L2  \citep{ExertierEtAl2014}, %{the single-photon detecor on ICESat-1 \citep{SunEtAl2004} } 
and those that are under development, such as {the single-photon} ELT \citep{ProchazkaEtAl2012}. %{and ICESat-2 \citep{AbdalatiEtAl2010} system}. 
Although the optical components of T2L2 and ELT are very different from those in an ILR system, the stable single-photon detection system has very similar characteristics. %{Detectors capable of accurate single-photon detection are also used in the altimeters of the ICESat-1 \citep{SunEtAl2004} and the upcoming ICESat-2 \citep{AbdalatiEtAl2010} missions.}

The measurement error due to the detector, often a Silicon Photon-Avalanche Diode (SPAD) is, at 3-6 mm, {the largest contributor to hardware-induced error budget} of SLR \citep{ExertierEtAl2006}. Space-grade detectors showing sub-ps stability have been developed for the ELT project \citep{ProchazkaEtAl2011,KodetEtAl2012}. %, providing confidence that space-based photon detectors are capable of operating at the same level of uncertainty as current ground-based systems.
{The contribution to the ILR precision of the best photon-counting detectors is typically 3 mm RMS \citep{ProchazkaEtAl2017b}. }

\cite{ExertierEtAl2006} give values of several mm for the influence of jitter in the event timer. A novel type of event timer, developed and applied by \cite{PanekEtAl2010,ProchazkaEtAl2011c}, provides sub-ps precision and a stability of several fs over a period of minutes to hours. The use of this technology allows the event timer to have an almost negligible contribution to the range error budget. 

For ILR, the influence of clock noise is substantially different from SLR/LLR \citep{Degnan2002,DirkxEtAl2015}. For two-way systems, the clocks only need to be {sufficiently stable} over short periods of time (two-way light time).  %For example, for LEO, lunar and planetary targets, which have light times on the order of 1 ms, 1 s and 1000 s, respectively, a clock with a relative stability of $10^{-14}$ (over these light times) will induce two-way range errors of 3 nm, 3 $\mu$m and 3 mm, respectively. 
A stability of about $10^{-15}$ over a typical ILR light time of 1000 s will result in 1 ps timing error (0.3 mm one-way range error), and is achievable by H-masers \citep[\emph{e.g.} ][]{DehantEtAl2017}. For the space segment, stability is only required over the time $\delta t$, %, see Eq. (\ref{eq:retransmissionTime})
putting sub-mm errors well within the capabilities of present spaceborne systems. A one-way range system requires clocks at both ends of the link to be stable over longer time periods \citep{BauerEtAl2016b}, making clock noise a significant issue.

Delays in various components of the electrical and optical system of the detection assembly must be accurately characterized to realize a high-quality range measurement. %In a typical SLR system, calibration is routinely performed. 
\cite{KirchnerEtAl2014} show that ground station calibration consistency on short time scales is at the several ps level {averaged over 10 s}, comparable to the value of 3 ps given by \cite{ProchazkaEtAl2012} for ELT.  Both are in line with the value of 1 mm given by \cite{ExertierEtAl2006}. Nevertheless, consistently obtaining mm-level system calibration on the space segment will be challenging.%, but a stable range bias can be largely removed during data analysis. %Similarly, \cite{ExertierEtAl2006} give state-of-the-art systematic uncertainties of 1 mm in both electronic and mechanical range uncertainties.

Data from existing two-way ILR experiments cannot be used to set up a measurement error budget.
{The only two-way ILR experiment thus far used the non-dedicated hardware on the MESSENGER spacecraft \citep{SmithEtAl2006}, which is not representative of the state-of-the-art.} Two-way links have been demonstrated on laboratory scales. \cite{ChenEtAl2013} %They use ultrashort (4 and 5 ps) laser pulses in a two-way setup over a distance of about 1.0 m, varying the distance by about 8 cm during the experiment. 
obtain range measurement errors below the 0.2 mm level (averaged over 1000 measurements). \cite{BlazejEtAl2014} have shown time transfer with an accuracy of 3 ps ($\approx$ 1 mm) using two representative ground segment hardware packages. These experiments show the capabilities of laboratory-scale experiments with well-controlled hardware, indicating the potential for (sub-) mm range accuracy.% (at least from a hardware point-of-view). %However, practical aspects related to system calibration over longer periods of time, comparison of distinct SLR and space segment systems, environmental instabilities, \textit{etc.} will likely make this infeasible in near-term ILR operations, limiting hardware-induced accuracy to the several-mm level.

%One-way laser ranging data to the LRO spacecraft are available over a period of five years \citep{MaoEtAl2016}. However, due to the one-way setup, the dominant noise source originates in the on-board and ground station clocks \citep{Bauer2017}, requiring the estimation of clock parameters. %as analyzed in detail by \cite{BauerEtAl2016b,BuccinoEtAl2016} using the LRO data and theoretically by \cite{DirkxEtAl2015}. 
%{Postfit residuals can be reduced to the measurement precision (10-20 cm), but doing so requires the estimation of a large number of clock parameters. These parameterscan absorb part of the signature of the spacecraft's orbital dynamics, complicating the interpretation of the results and potentially reducing orbit quality \citep{BauerEtAl2016b,DirkxEtAl2016}.}
%The remaining clock noise in the data remains at the level of $>1$ m \citep{DirkxEtAl2016}, the consequences of which are %at least when limiting the number of clock parameters that are estimated, 
%discussed in more detail in Section \ref{sec:oneTwoWayScience}. %When estimating many clock parameters, correlations with the state estimation degrade the true error of the final orbit determination. This makes one-way ranging a much less attractive option (see Section \ref{sec:oneTwoWayScience}). %As an \emph{ad hoc} product, laser communications data from the LADEE spacecraft was used to obtain two-way ranging data with a precision of several cm \citep{StevensEtAl2016}. 

\subsection{Data Analysis Uncertainties}
\label{sec:modelErrorSources}

Even in the case of perfect range measurements ($\varepsilon_{s}=0)$, the science products obtained from the data will not be error-free. %, due to inaccuracies in the models used to map the observations to the estimated parameters. %This can result in a signal of a parameter $p_{1}$ being partly misattributed to a parameter $p_{2}$, skewing the estimation of both parameters. Also, it can limit the model from fully fitting the trend in the measurement residuals to the parameterized model, preventing the data from being fully exploited.
Errors in the evaluation of Eq. (\ref{eq:prelimRange2}) %(the state function of the link ends $\mathbf{r}_{A}(t)$ and $\mathbf{r}_{B}(t)$, and corrections $\Delta s_{BA}$) 
will degrade the fidelity of the results. %For SLR and LLR, the errors in the various components of the range calculation primarily stem from the following issues:

The position of the ground station in the Geocentric Celestial Reference System (GCRS) requires a time-dependent position in the International Terrestrial Reference Frame (ITRF), as well as a rotation between the two.  
%Firstly, a model for the (time-dependent) position of the ground station reference point consists of a position in an Earth-fixed frame, such as the International Terrestrial Reference Frame (ITRF) as a function of time and a rotation from this frame to the Geocentric Celestial Reference System \citep[GCRS; ][]{PetitEtAl2010}. 
Inaccuracies in these models %, including uncertainties in Earth deformation/rotation, 
limit the accuracy of the ground station position function at the sub-cm level \citep{AltamimiEtAl2011,Rothacher2011, SosnicaEtAl2013}. %Additionally, inaccuracies of Earth deformation models, both tidal and non-tidal, limit the accuracy of ground station position models \citep{SosnicaEtAl2013} to the several mm level.

ILR analysis must be done in the Barycentric Celestial Reference System (BCRS). As a result, uncertainties in the \textit{a priori} Earth ephe\-meris will enter the error budget of the ground station position model. % for ILR, since the analysis should be done w.r.t. the solar system barycenter. 
The ephemeris of the Earth is orders of magnitude less accurate than the expected cm-accuracy of ILR measurements \citep{FiengaEtAl2011}. This indicates that the Earth's ephemeris {should} be among the estimated parameters during ILR data analysis. %  to prevent its \textit{a priori} errors from reducing the analysis quality. 

%The state function of the space segment (lander, orbiter) is typically fitted to some parametric model during the estimation process. However, a variety of effects prevent the space segment dynamics from being modelled to the inherent ILR measurement accuracy over long time intervals. %In particular, mismodelling of the non-conservative forces acting on the spacecraft is a persistent issue. 

For landers on solar system bodies, the general issues in modelling the bary\-centric state are of a similar nature as for Earth ground stations. %: uncertainties in rotation, tidal deformation and ephe\-meris of the body. 
Depending on the target body, however, the uncertainty may be limited and accounted for by the addition of a number of estimated parameters. In fact, the signatures of these effects will often be key science objectives of the lander tracking \citep[\emph{e.g.} ][]{LeMaistreEtAl2013,DirkxEtAl2014}. Fulfilling the modelling requirements may require significant theoretical work. For planetary dynamics, the uncertainty in asteroid masses and orbits is presently the limiting factor in the dynamical models \citep{FiengaEtAl2009}. For the dynamics of natural satellite systems, the consistent coupling between translational and rotational dynamics and tidal deformation will be challenging to model at the mm-level \citep[\emph{e.g.} ][]{DirkxEtAl2016}.

For orbiters, uncertainties in non-conservative forces, as well as target body gravity field variations \citep[\emph{e.g. }][]{MartyEtAl2009}, can limit the accuracy to which the dynamics can be modelled. This requires the state estimation to be performed arc-wise (a typical arc length is several days). %Additionally, uncertainties in a body's time-varying gravitational field can diminish the attainable accuracy of the estimated orbit and parameters \citep{MartyEtAl2009}. %unless significant supplemental data is available. In fact, 
Dynamical model error can become the dominant source of uncertainty in the estimated parameters, and is a key reason for the true estimation errors often being significantly higher than the formal estimation errors \citep[\emph{e.g.} ][]{KonoplivEtAl2011,MazaricoEtAl2014b}.

Finally, %uncertainty in the term $\Delta s_{BA}$ also enters the error budget of ILR data analysis. 
models for tropospheric correction (as part of $\Delta s_{BA}$) have {an accuracy of} 5-8 mm \citep{ExertierEtAl2006}. This level can be reduced to the (several) mm level using ray-tracing models \citep{HulleyPavlis2007}. %Similar to the analysis of radio tracking data, it is the uncertainty in the wet contribution of the tropospheric delays that causes the majority of the error. %Relativistic range corrections for ILR will require the consideration of various effects that have not heretofore been included in the analysis of tracking data. However, 
Detailed models for relativistic range corrections  have been developed \citep[\emph{e.g.} ][]{TeyssandierPoncinLafitte2008}, so model uncertainties for this contribution of $\Delta s_{BA}$ will be negligible {if} state-of-the-art models are applied.

\subsection{Total Uncertainty - Summary}
\label{sec:ilrTotalError}
In Sections \ref{sec:inherentStochastic}-\ref{sec:modelErrorSources}, we have presented the main error sources that enter the data realization and analysis chain of ILR. {The main sources of measurement error are the detector uncertainty (at 3 mm), and the finite pulse length (3 to 13 mm RMS, for laser pulses 10 to 100 ps FWHM respectively).}

 {%The resulting ILR precision is in a range of 3 to 13 mm RMS, for laser pulses 10 to 100 ps long respectively.  %At the single-photon level, several measurements (typically > 10) have to be collected in one series to be able to identify the correct signal. 
As a result, 1.0 to 4.3 mm precision averaged over 10 measurements (for 10 to 100 ps pulse length) may be achieved.
%Combining single-shot data points into a normal point allows these random effects to be brought to $<1$ mm precision.
Considering the existing detector and timing devices performance (Section \ref{eq:measErrors}), a limiting precision (but not accuracy) $<$0.1 mm can be achieved when averaging higher number of individual single photon measurements.  }

Hardware imperfections, as they are deduced from current SLR and space-based laser transmitter and detector systems, {will induce systematic errors at the several mm level,  as is the case for SLR.} %In the near future, although sub-mm accuracy has been demonstrated at laboratory scales. %System-induced uncertainties will likely prevent mm-accurate ILR data from being consistently achieved, at least at the present time. 

It is especially the instabilities in the systems that will be an issue for the quality of ILR data (as well as for SLR/LLR). A system bias that is constant {for an extended time interval} can be mitigated by the estimation of {a single} long-arc range bias. Instabilities (\emph{e.g.} random walk behaviour) cannot be removed in this manner without introducing an excessive number of parameters. Therefore, ILR data accuracy will be limited to the level of several mm at best. Considering the current performance of SLR systems, sub-cm accuracy will be feasible. A major design driver for the space segment will be the stable system delay calibration in an (inter)planetary environment. Existing model errors in tropospheric correction and ground station positioning will limit ILR data modelling to the several mm level  (Section \ref{sec:modelErrorSources}). These issues are also crucial in space geodesy, and are a topic of active research. 

Dynamical model errors of both the space segment and the Earth will limit the accuracy to which the data can be interpreted. The degree to which these errors will affect the estimation of parameters of interest is strongly dependent on the correlation of the signatures of these parameters with the model errors. This error source is similar for both Doppler and range data, and can prevent a data set from being exploited to its full potential.

\section{Data-Type Comparison Methodology}
\label{sec:compCrit}
In this section, we present the methods we use to compare ILR with radio data. We outline our concept for analytical comparison and numerical covariance analysis in Sections \ref{sec:snr} (based on \cite{Dirkx2015}), and Section \ref{eq:covarianceSettings}, respectively. %Our numerical approach provides a validation for the analytical method, serves to identify the limits of its applicability and acts as a guide for the interpretation of the analytical results.

\subsection{Analytical Approach}
\label{sec:snr}
%To compare the range and range-rate observables, it is to be evaluated how their total measurement value, and time-variations of these values, relate to the influence of a parameter $q$ that is to be estimated. In this section we derive a criterion based upon which the contribution  of the range and range-rate observables to the estimation of physical parameters can be quantitatively compared. 

The sensitivity of an observable $h$ to a parameter $q$ is quantified by its associated partial derivative $\partial h/\partial q$, \citep[\textit{e.g.}, ][]{MontenbruckGill2000}. Since the range observable $s$ and range-rate observable $\dot{s}$ are related through Eq. (\ref{eq:prelimRangeRate}), we have:
\begin{align}
\frac{\partial\dot{s}(t)}{\partial q}=\frac{1}{\Delta t_{i}}\left(\frac{\partial {s}(t+\Delta t_{i})}{\partial q}-\frac{\partial {s}(t)}{\partial q}\right)\label{eq:rangeRatePartial}
\end{align}
%This relation shows that  the range-rate observable will be completely insensitive to parameters $q$ which have a constant influence on the range measurement ($\partial s/\partial q=$ constant). Similarly, the range rate observable will be only weakly sensitive to periodic signals of a parameter $q$ for which the period $T$ is much larger than $\Delta t_{i}$. 

To quantitatively compare the data types, we define
%a criterion that describes the influence of a unit change in $q$ on bothn range and range-rate by comparing it to the noise levels of $s$ and $\dot{s}$ (which we denote $\sigma_{s}$ and $\sigma_{\dot{s}}$, respectively). Essentially, this criterion is
a signal-to-noise (SNR) criterion for an observable $h$ and a parameter $q$, denoted SNR$_{h;q}$, which is computed as follows: 
%which represents the magnitude of the observable signal (from Eq. (\ref{eq:rangeRatePartial}) for range) w.r.t. the noise level of the data. For any observable $h$, the SNR$_{h;q}$ of a parameter $q$ is then defined as:
\begin{align}
\text{SNR}_{h;q}=\left|\frac{1}{\sigma_{h}}\frac{\partial h}{\partial q}\right|\label{eq:observableSnr}
\end{align}
%which is a first-order estimate of the relative sensitivity of an observable $h$ to a parameter set $q$. % by assuming that the observable value $h$ is composed of discrete increments of $\sigma_{h}$, and omitting any correlations with other parameters.
where $\sigma_{h}$ is the noise level of the measurement $h$.

Now, we define the following figure of merit to compare the relative sensitivity of range and Doppler observables to a parameter $q$:
\begin{align}
\Xi_{q}=\frac{\max_{t}\left(\text{SNR}_{\dot{s};q}\right)}{\max_{t}\left(\text{SNR}_{{s};q}\right)}\label{eq:rangeRangeRateCompareCriterion}
%\frac{\partial s}{\partial q}|_{t}&>\frac{\sigma_{s}}{(\Delta t_{i})\sigma_{\dot{s}}}\left(\frac{\partial s}{\partial q}|_{t+\Delta t_{i}}-\frac{\partial s}{\partial q}|_{t}\right)
\end{align}
%Using this criterion, the qualitative comparative discussion of range and range rate in Section \ref{sec:preliminaryTypeComparison} can be quantified, which we present and discuss for a simplified behaviour of the parameter $q$ in Section \ref{sec:periodicSignalAnalysis}. 
where the maximum is taken over the observational period. To first order, we can set $\Xi_{q}<1$ as a criterion when ILR would become a {feasible} alternative to Doppler data for determining a parameter $q$. The interpretation of numerical values of this criterion is discussed in Section \ref{sec:observationUncertaintyComparison}. 

We start by using an analytical model for Eq. (\ref{eq:rangeRatePartial}), in which the influence of a parameter $q$ is manifested in the range measurements as a purely sinusoidal signal of amplitude $A$ and angular frequency $\omega$ (period denoted as $T$), so that:
\begin{align}
\frac{\partial s}{\partial q}&=A\sin(\omega t) \label{eq:rangeToyModel}
\end{align}
and we obtain the following from Eq. (\ref{eq:rangeRatePartial}):
\begin{align}
\frac{\partial\dot{s}(t)}{\partial q}&=\frac{A}{\Delta t_{i}}\Big( \big( \cos(\omega \Delta t_{i})-1\big)\sin(\omega t) +\sin(\omega \Delta t_{i})\cos(\omega t ) \Big) \label{eq:rangeRateToyModel2}
\end{align}
%
%\begin{align}
%\frac{\partial\dot{s}(t)}{\partial q}&=\frac{1}{\Delta t_{i}}\left(\frac{\partial {s}(t+\Delta t_{i})}{\partial q}-\frac{\partial {s}(t)}{\partial q}\right)\\
%&=\frac{1}{\Delta t_{i}}\left(A\sin(\omega(t+\Delta t))-A\sin(\omega(t))\right)\\
%&=\frac{A}{\Delta t_{i}}\left(\cos(\omega\Delta t_{i})\sin(t)+\sin(\omega\Delta t_{i})\cos(t)-\sin(\omega(t))\right)
%\end{align}

This approximates the situation where all planets are in circular orbits around a static Sun, with the spacecraft in a circular orbit around one of these bodies, and the parameter $q$ imparting an $N$-cycles-per-orbit sinusoidal signal on the data (with $N$ an integer). In the data, the orbital frequencies of the spacecraft, Earth and the target planet will then all be visible\footnote{For various parameters, the sinusoidal signature will be modulated by a linear trend, increasing the amplitude with time. However, if the observation time is much larger than the period of the signal $T$, the impact of this linear trend on Eq. (\ref{eq:rangeRangeRateCompareCriterion}) will be small.}. %since the trend linear trend will not change the relative strength of the observables.}. 
These assumptions are a reasonable approximation {for our analysis}, {as validated in Section \ref{eq:numericalResults} from numerical results}. In Appendix \ref{ref:semiAnalyticalResults}, we discuss how to extend the method to elliptical orbits.

%Using these assumptions, the sensitivity to the parameter $q$ becomes:
%\begin{align}
%\frac{\partial s}{\partial q}&=A\sin(\omega t) \label{eq:rangeToyModel}\\
%\frac{\partial \dot{s}}{\partial q}%&=\frac{1}{\Delta t_{i}}\Big(A\sin\big(\omega (t+\Delta t_{i})\big)-A\sin(\omega t)\Big) \label{eq:rangeRateToyModel}\\
%&=\frac{A}{\Delta t_{i}}\Big( \big( \cos(\omega \Delta t_{i})-1\big)\sin(\omega t) +\sin(\omega \Delta t_{i})\cos(\omega t ) \Big) \label{eq:rangeRateToyModel2}
%\end{align}
For $\Delta t\ll T$, we obtain the following from Eq. (\ref{eq:rangeRateToyModel2}):
\begin{align}
\lim_{\omega\Delta t_{i}\rightarrow 0}\left(\max_{t}\left(\frac{\partial \dot{s}}{\partial q}\right) \right)=\frac{2\pi A}{T}=\omega A
\end{align}
so that:
\begin{align}
\Xi_{q}\biggr\rvert_{\Delta t\ll T}\approx\frac{\sigma_{{s}}}{\sigma_{\dot{s}}}\omega\label{eq:approximateAnalyticalRatio}
\end{align}
{We apply this limit approximation in Sections \ref{sec:results} and \ref{sec:discussion} to compare the analytical and numerical approaches.}

\subsection{Covariance Analysis - Numerical Models}
\label{eq:covarianceSettings}
To assess the validity of our analytical results, and gain insight into how the results obtained from them should be interpreted, we perform covariance analyses \citep[\emph{e.g. }][]{MontenbruckGill2000,MilaniEtAl2010} for a number of representative cases. We generate formal errors for a set of parameters when using only Doppler data, and when using only range data. {We denote the formal error of parameter $p$, using data type $h$, as $\epsilon_{p,h}$}.\footnote{Note that we use the symbols $\varepsilon$ (lumped range error) and $\epsilon$ to represent
different physical quantities.}%, to assess the influence of a parameter's characteristic frequency on the ratio in formal errors.

We simulate two scenarios: a Mars lander mission, and a Mars dual-orbiter mission. The orbits of the spacecraft are both low-altitude, nearly polar and nearly circular, with initial condition $e=0.01$ for both; $a=3850$ km and $i=88^{\circ}$ for one orbiter and $i=92^{\circ}$ and $a=3800$ km for the other orbiter (similar to spacecraft such as Mars Odyssey and Mars Reconaissance Orbiter). The lander is placed equatorially. For both cases, we assume a simplified Mars rotation model, with fixed pole right ascension and declination $\alpha_{M}$ and $\delta_{M}$, and a fixed rotation rate $\omega_{M}$. %In the orbiter simulations, we restrict ourselves to the near-circular orbit case.

For the orbiter simulation, we estimate the initial states $\mathbf{x}_{i}$  of both orbiters ($i=1,2$) w.r.t. Mars' center of mass over 1-week arcs, with each arc $j$ starting at time $t_{j}$ over a 2-year period ($j=1..104$). For both the lander and the orbiter simulations, we estimate Mars' dynamics over a single 2-year arc w.r.t. the barycenter $\mathbf{x}_{M}(t_{0})$ {(see Table \ref{tab:typicalMissionsPositionVelocityCompare}).}. 

We set up our state transition matrix $\Phi(t,t_{0})={\partial\mathbf{x}}/{\partial\mathbf{x}(t_{0})}$ (with $\mathbf{x}=[\mathbf{x}_{M}; \mathbf{x}_{1}(t_{1});...;\mathbf{x}_{1}(t_{104}); \mathbf{x}_{2}(t_{1});...;\mathbf{x}_{2}(t_{104}) ]$) in such a way that the coupling terms $\partial\mathbf{x}_{i}(t)/\partial\mathbf{x}_{M}(t_{0})$ are referenced to the correct arc, so that for a given arc $j$:
\begin{align}
\frac{ \partial\mathbf{x}_{i}(t)}{\partial\mathbf{x}_{M}(t_{0})}=\delta_{ij}\frac{ \partial\mathbf{x}_{i}(t)}{\partial\mathbf{x}_{i}(t_{j})}\frac{ \partial\mathbf{x}_{i}(t_{j})}{\partial\mathbf{x}_{M}(t_{0})},\,\,  t \in [t_{j}, t_{j+1}) \label{eq:hybridStateTransitionMatrix}
\end{align}
{with $\delta_{ij}$ the Kronecker delta function. By doing so, the estimation of spacecraft state and natural body state can be done concurrently, as the influence of a change of natural body state is directly mapped to a change in (barycentric) spacecraft state.}

This approach is in contrast to the typical approach of orbit determination and ephemeris generation, where the spacecraft's orbit is first estimated using Doppler data only, and the planetary ephemerides are then estimated using range/VLBI data (without adjusting the spacecraft orbit). In {this traditional approach, the} direct coupling between the planetary and spacecraft orbits is omitted. When incorporating ILR, however, the laser data will have significant and useful information on the dynamics of both natural and artificial bodies, requiring the coupling to be incorporated into the simulations. {Except for the modification in the computation of $\Phi(t,t_{0})$ as in Eq. (\ref{eq:hybridStateTransitionMatrix}), our covariance analysis follows that outlined by, \emph{e.g.}, \cite{MontenbruckGill2000}}.

\begin{table*}[tbp]
\centering
\small
\caption{Estimated parameters for the numerical simulation cases. }
\begin{tabular}{l | c c c c c c c }
\hline
Simulation case & $\mathbf{x}_{i}(t_{j})$ & $\mathbf{x}_{M}(t_{0})$ & $\mathbf{x}_{L}$ & $\omega_{M}$, $\alpha_{M}$, $\delta_{M}$ & $(C,S)_{l,m}$, $k_{20}..k_{22}$ & $\beta$, $J_{2_{\astrosun}}$ \\
\hline
\hline
%Orbiters A & \checkmark & & & & &\\
%Orbiters B & \checkmark & \checkmark & & & & \\
Orbiters & \checkmark & \checkmark & & \checkmark & \checkmark &\\
%Lander & & \checkmark & &  &  & &\\
%Lander & & \checkmark & \checkmark&  \checkmark &  &\\
Lander & & \checkmark & \checkmark & \checkmark &  & \checkmark \\
\hline
\end{tabular}
\label{tab:typicalMissionsPositionVelocityCompare}
\end{table*} 

The list of parameters we estimate for both scenarios is given in Table \ref{tab:typicalMissionsPositionVelocityCompare}. {The relevance of the parameters in the context of Mars missions is discussed by \emph{e.g}, \cite{ KonoplivEtAl2011, RivoldiniEtAl2011})}. %The symbols $\omega_{M}$, $\alpha_{M}$ and $\delta_{M}$ denote the constant rotation rate, pole right ascension and pole declination of Mars. 
The notation $(C,S)_{l,m}$ is used as a shorthand for the combination of spherical harmonic coefficients $C_{lm}$ and $S_{lm}$ of Mars.  For the Sun, we consider only the degree-two zonal coefficient $J_{2,\astrosun}$. The Love numbers are denoted as $k_{lm}$, and $\beta$ denotes the PPN parameter \citep{Will2014}. 
%Although we focus on geodetic observables we also include the parametric post-Newtoian (PPN) parameter $\beta$, which is key parameter in experimental relativity \citep{Will2014}. %We generate covariance matrices for several parameter vectors $\mathbf{q}$ (indicated as sets A, B and C), starting at a basic set (initial states only), extending it to more elaborate and realistic sets (gravity field coefficients, rotational parameters, \emph{etc.}). 
Our simulations represent a reduced analysis, not a full mission analysis. Instead, it is geared towards investigating the contribution of the range and Doppler data types to the estimation of various parameters, and comparing the results to those obtained from Section \ref{sec:snr}.%, and ascertaining the degree of validity of the simplified model discussed in Section \ref{sec:compCrit}.

\begin{figure*}[tb!]
\centering
\includegraphics[width=0.77\textwidth]{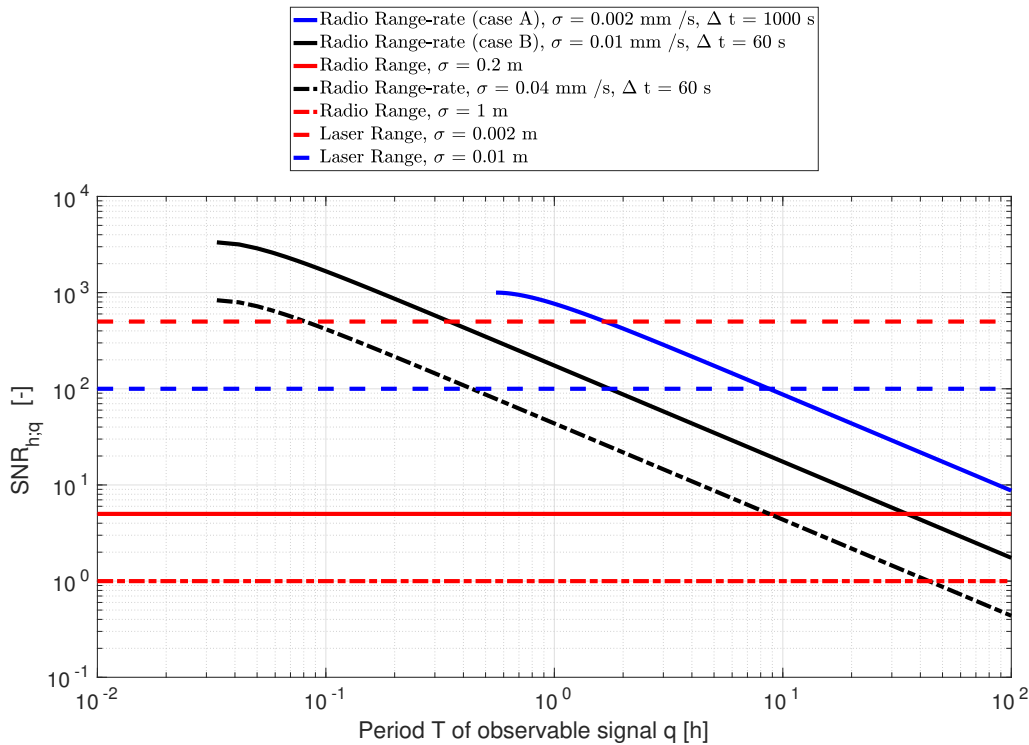}
\caption{Comparison of range and range-rate observable to purely periodic signals of amplitude $\omega$. Dashed line represents ILR, thick lines next-generation radiometric, dashed-dotted current radiometric. Adapted from \cite{Dirkx2015}.}
\label{fig:rangeRangeRateCompre}
\end{figure*}

For both data types, we use a daily tracking pass of 2 hour duration, generating one independent measurement every 60 s. The laser ranging data is weighted at 1 cm, and the Doppler data at 0.01 mm/s (see Section \ref{sec:observationUncertaintyComparison} for more detailed discussion). We do not simulate data for a solar separation angle smaller than 5$^{\circ}$. %We do not discuss the details of the covariance analysis procedure here, models can be found in \citep[\emph{e.g. }][]{MontenbruckGill2000,MilaniEtAl2010}. 
For our simulations, we use the Tudat software\footnote{Documentation at \url{http://tudat.tudelft.nl}; Code at \url{http://github.com/tudat/tudatBundle}}.

\section{Results}
\label{sec:results}

Using the methods outlined in Section \ref{sec:compCrit}, we give the results of our analytical and numerical analyses. %comparison (Section \ref{sec:analyticalResults}), semi-analytical comparison (Section \ref{ref:semiAnalyticalResults}) and numerical comparison (Section \ref{eq:numericalResults}) of Doppler and ILR data types. 
%For the analytical comparison, we limit ourselves to the comparison of the partial derivatives (Section \ref{sec:snr}). For the numerical approach, we compare the formal errors (Section \ref{eq:covarianceSettings}).

%With the criterion defined by Eqs. (\ref{eq:observableSnr}) and (\ref{eq:rangeRangeRateCompareCriterion}), we are in a position to make a conceptual comparison between the capabilities of an ILR system and a radiometric Doppler system of representative current and near-future capabilities. 

%Using representative values of the laser range data quality (Section \ref{sec:ilrTotalError}) and radiometric data data quality and integration time given (Section \ref{sec:radiometricTrackingDataQuality}), we use Eq. (\ref{eq:observableSnr}) to evaluate the sensitivities of these observables to periodic signals. To compare the performance of range and range-rate systems, we calculate  the maximum values of Eqs. (\ref{eq:rangeToyModel}) and (\ref{eq:rangeRateToyModel2}) over a single period to determine the maximum observable signal of the parameter $q$ on both $s$ and $\dot{s}$, respectively.  

%We use Eq. (\ref{eq:rangeRangeRateCompareCriterion}) to determine for which periods $\omega$ of the signal $q$ the range measurements become superior to Doppler measurement (in terms of SNR$_{h;q}$), for our assumption of purely periodic signals.

\subsection{Analytical Results}
\label{sec:analyticalResults}

Here, we present the results of the analysis method described in Section \ref{sec:snr}. For current radiometric systems, we use accuracies of 1 m in range and 0.04 mm/s in range rate for $\Delta t_{i}=$60 s. For upcoming state-of-the-art systems, we assume 0.2 m accuracy in radiometric range, and both 0.01 mm/s at 60 s integration time and 0.002 mm/s at 1000 s integration time ({Section \ref{sec:radiometricTrackingDataQuality}}). For ILR, we assume an upper and lower bound on data quality of 2 and 10 mm, respectively, {a broad range which we derive from our discussion in} Section \ref{sec:ilrTotalError}. Evaluating Eqs. (\ref{eq:rangeToyModel}) and (\ref{eq:rangeRateToyModel2}) to approximate the behaviour of the partial derivatives gives the results for the SNR$_{h,q}$ from Eq. (\ref{eq:observableSnr}), shown in Fig. \ref{fig:rangeRangeRateCompre}. %There, the quantity SNR$_{h,q}$ for these data types to periodic signals are shown as a function of the period $T$ of the signature imparted by the parameter $q$. 	

%If laser ranging systems are to be implemented in the future, it should be shown that it has a clear scientific advantage over next-generation Doppler systems.
%The range-rate noise level decreases with integration time \citep{AsmarEtAl2005} under typical conditions, so that a range-rate observation with a larger $\Delta t_{i}$ inherently has a better SNR.  
%For cases where $T<\Delta t_{i}$, however, a $2\pi$ ambiguity arises, requiring the use of a smaller integration time. 
%Range measurements could be used for the estimation of such high-periodic signals, but with a much reduced sensitivity compared to short integration time range-rate measurements, since $\frac{\partial s}{\partial q}$ does not increase with decreasing $T$, as does $\frac{\partial \dot{s}}{\partial q}$. 

The figure shows that the dual-frequency Doppler data at $\Delta t_{i}=$60 s ($\sigma$=0.01 mm/s) and the ILR curves cross in the area of 0.33-1.65 hours (for range data precision of 2-10 mm), and therefore $\Xi_{q}<1$ for lower $T$ and $\Xi_{q}>1$ for larger $T$. This time interval is a particularly interesting one, as it contains the orbital period of many spacecraft orbiting rocky/icy bodies. For the $\Delta t_{i}=$1000 s Doppler data, this time interval is shifted by a factor {of 5 (as the 1000 s data has a precision 5 times better than the 60 s data).} %which we obtain from Eq. (\ref{eq:approximateAnalyticalRatio}) and the fact that the range-rate uncertainty is 5 times lower for the 1000 s integration time, compared to the 60 s value}. 
For cases where $T<\Delta t_{i}$, however, a $2\pi$ ambiguity arises in the estimation, requiring the use of a smaller $\Delta t_{i}$. Doppler data for orbiters typically have an integration time of 60 s or smaller. Reducing the integration time of the data does result in greater data volume, reducing the formal estimation errors if the noise is not correlated in time. %The range-rate noise level decreases with integration time \citep{AsmarEtAl2005} under typical conditions, so that a range-rate observation with a larger $\Delta t_{i}$ inherently has a better SNR.  For cases where $T<\Delta t_{i}$, however, a $2\pi$ ambiguity arises, requiring the use of a smaller integration time. 

\begin{figure}[tb!]
\centering
\includegraphics[width=0.49\textwidth]{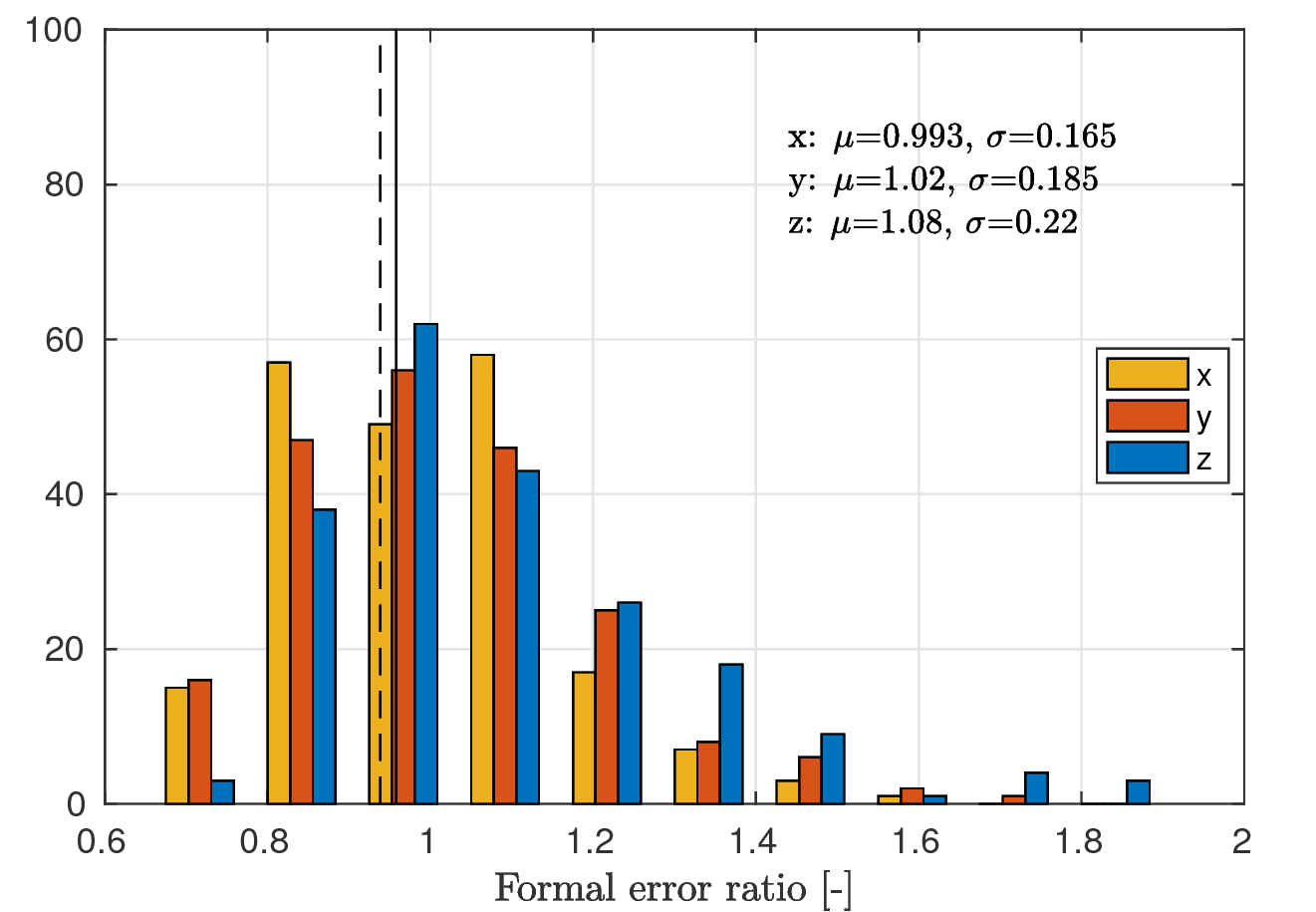}
\caption{Histograms of formal error ratio $\epsilon_{s}/\epsilon_{\dot{s}}$ for all arcs of the orbiter position component estimation, generated using the settings described in Section \ref{eq:covarianceSettings}. Vertical black lines represent the analytical ratios $\Xi_{q}$ at the spacecraft orbital periods.}
\label{fig:orbiterErrorRatio}
\end{figure}

\begin{figure*}	
	\centering
	\begin{subfigure}[t]{.47\textwidth}
		\centering
		\includegraphics[width=0.95\textwidth]{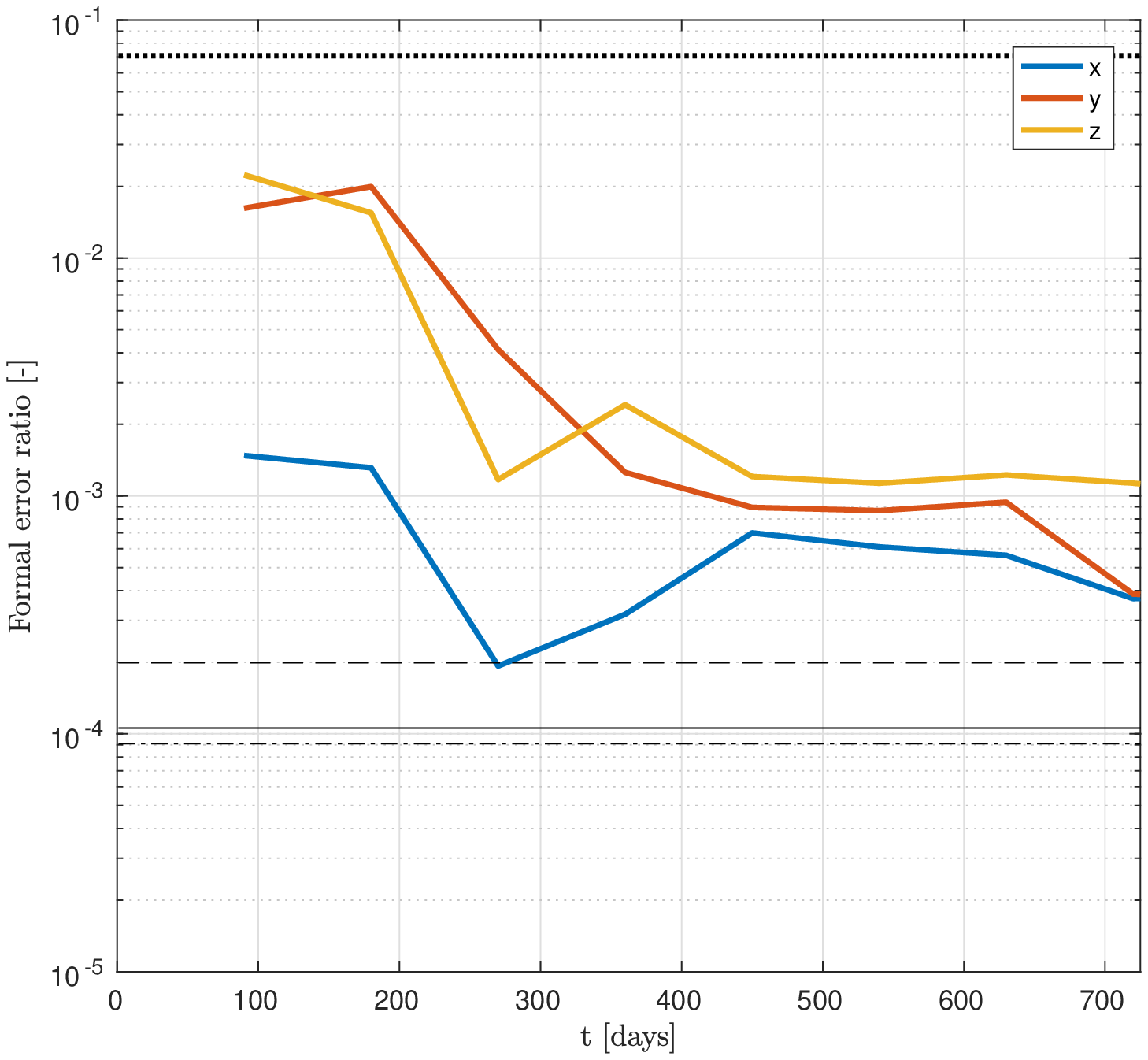}
		\caption{Orbiter simulation}
		\label{fig:orbiterMarsErrorRatio}		
	\end{subfigure}
	\quad
	\begin{subfigure}[t]{.47\textwidth}
		\centering
		\includegraphics[width=0.97\textwidth]{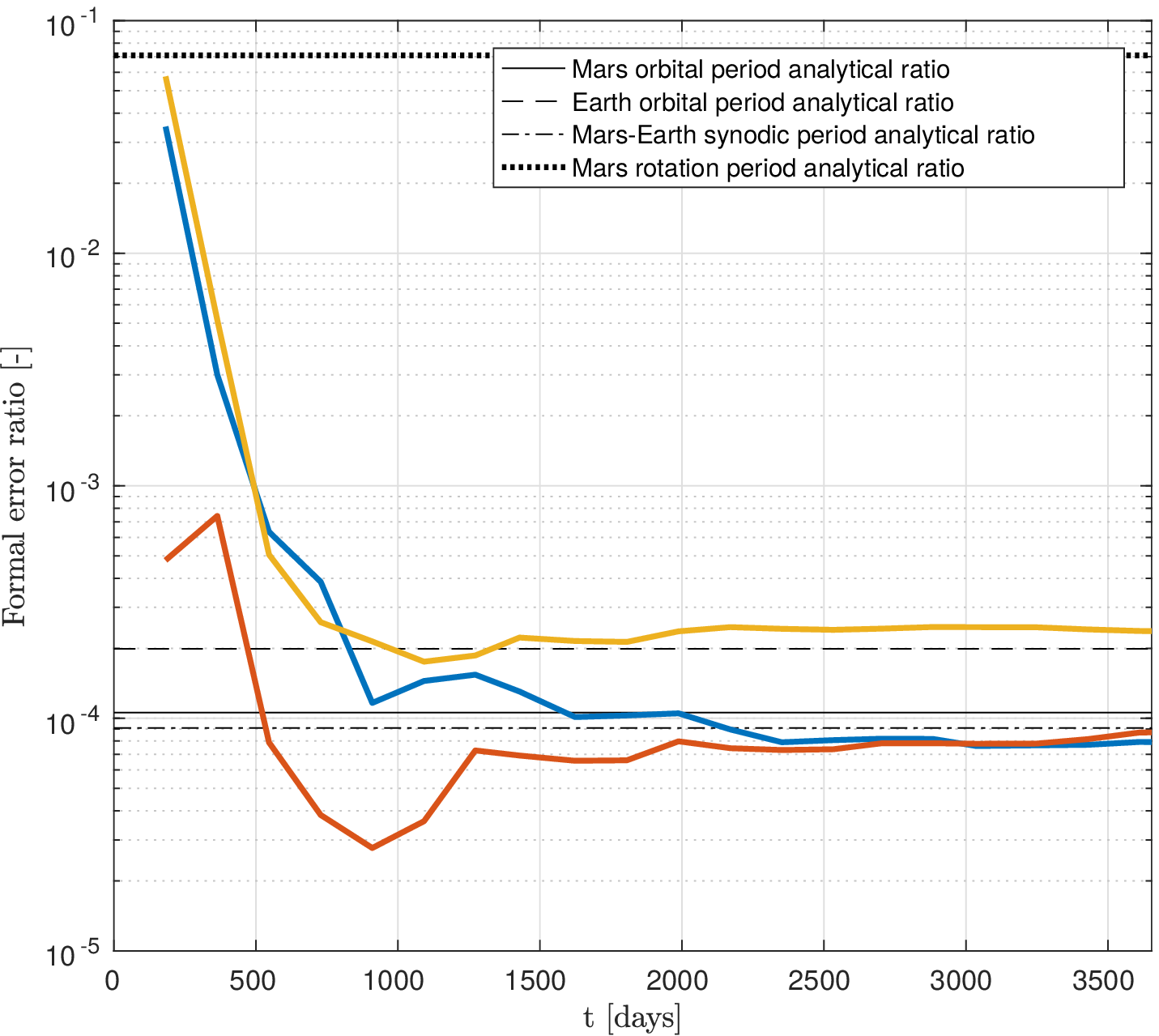}
		\caption{Lander simulation}
		\label{fig:landerMarsErrorRatio}
	\end{subfigure}
\caption{Formal error ratios $\epsilon_{s}/\epsilon_{\dot{s}}$ for the Mars initial position components as a function of mission duration, from the settings described in Section \ref{eq:covarianceSettings}. Horizontal black lines in a) represent analytical ratios $\Xi_{q}$ at relevant periods.}
\label{fig:marsErrorRatio}
\end{figure*}

\begin{figure*}
\centering
\begin{subfigure}{.49\textwidth}
\includegraphics[width=0.97\textwidth]{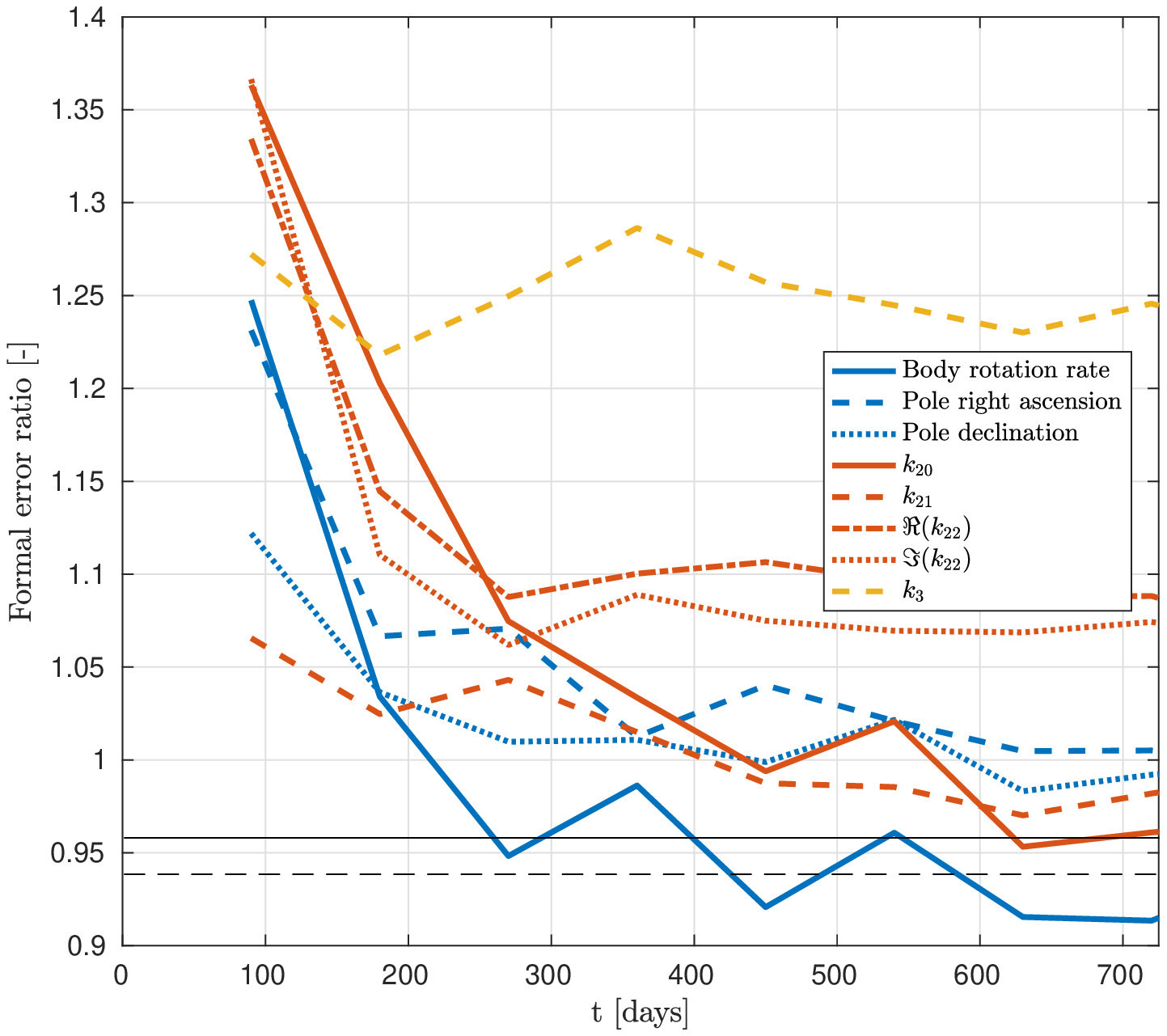}
\caption{Orbiter simulation}
\label{fig:orbiterParameterErrorRatio}
\end{subfigure}
\begin{subfigure}{.49\textwidth}
\includegraphics[width=0.95\textwidth]{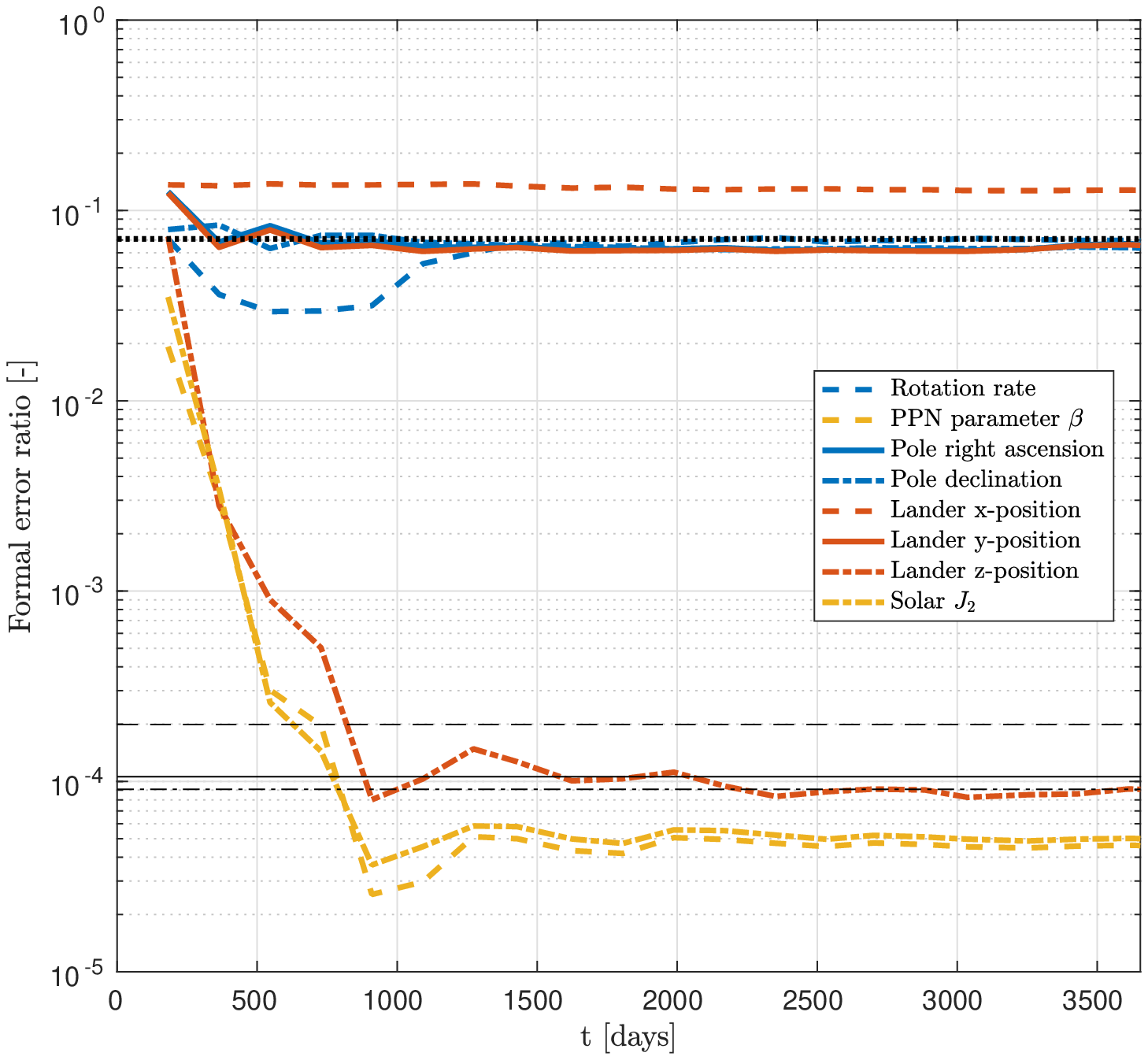}
\caption{Lander simulation}
\label{fig:landerParameterErrorRatio}
\end{subfigure}
\caption{Formal error ratios $\epsilon_{s}/\epsilon_{\dot{s}}$ for estimated parameters  as a function of mission duration, from the settings described in Section \ref{eq:covarianceSettings}. Horizontal black lines in a) represent the analytical ratios $\Xi_{q}$ at the spacecraft orbital periods. {Horizontal black lines in b) represent analytical ratios $\Xi_{q}$ at relevant periods, as in Fig. \ref{fig:marsErrorRatio}.}}
\label{fig:parameterErrorRatio}
\end{figure*}

\begin{figure}[tb!]
\centering
\includegraphics[width=0.49\textwidth]{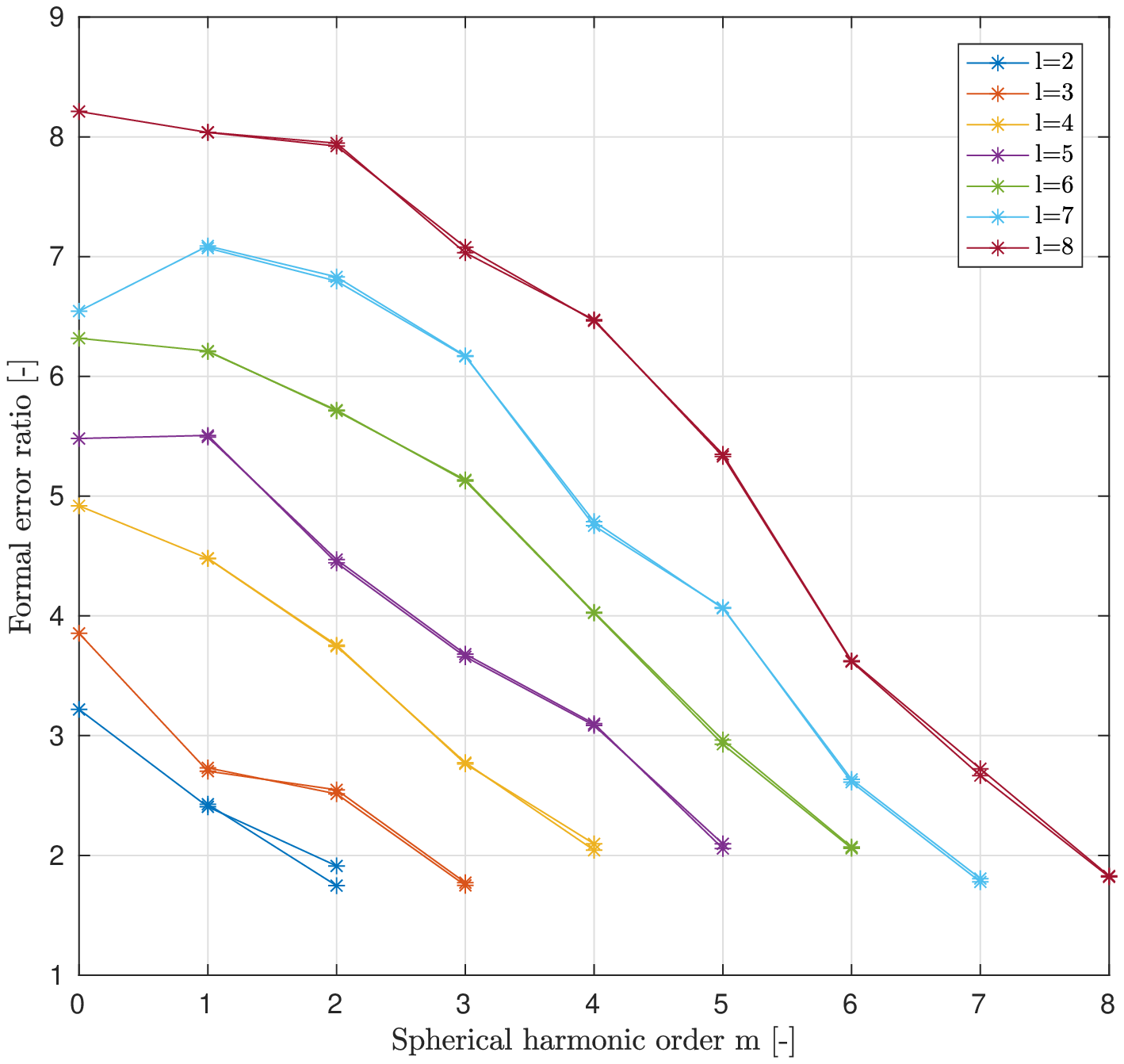}
\caption{Formal error ratios $\epsilon_{s}/\epsilon_{\dot{s}}$ for gravity field coefficients at $t=$2 years from the settings described in Section \ref{eq:covarianceSettings}.}
\label{fig:orbiterNumericalGravityFieldError}
\end{figure}

\subsection{Numerical Results}
\label{eq:numericalResults}
{In this section, we show the results of the covariance analyses described in Section \ref{eq:covarianceSettings}. %We analyze both a Mars orbital and Mars lander, with properties defined in the second paragraph of Section \ref{eq:covarianceSettings}. The estimated parameters for the two cases are listed in Table \ref{tab:typicalMissionsPositionVelocityCompare}}.
These results are} shown in Figs. \ref{fig:orbiterErrorRatio}-\ref{fig:orbiterNumericalGravityFieldError}. For these results, we use the formal error ratio $\epsilon_{q,s}/\epsilon_{q,\dot{s}}$ as figure of merit, instead of $\Xi_{q}$ {from Eq. (\ref{eq:approximateAnalyticalRatio}).}
Formal errors are a more robust indicator of the relative strength of the observables, as it includes the difference in correlations when performing Doppler-only and range-only estimation. {In the analytical analysis of Section \ref{sec:analyticalResults}, formal errors cannot be obtained, as no estimation is performed. In the remainder of this section, we analyze how well the two figures of merit compare to one another. The analytical approach is a reasonable approximation if $\Xi_{q}\approx\epsilon_{q,s}/\epsilon_{q,\dot{s}}$}.

The orbiter arc initial position error ratios $\epsilon_{r_{i,0},s}/$ $\epsilon_{r_{i,0},\dot{s}}$ are  shown in Fig. \ref{fig:orbiterErrorRatio}. The mean ratio ($\approx1.0$) is slightly larger than the theoretical ratio $\Xi_{q}$ ($\approx 0.95$, see Fig. \ref{fig:rangeRangeRateCompre}, with settings from Section \ref{eq:covarianceSettings}). Deviations from the analytical value are primarily due to the correlations between the various parameters, which are slightly worse for the range-only case than for the Doppler-only case.

{The ratio $\epsilon_{r_{M},s}/\epsilon_{r_{M},\dot{s}}$ (Mars initial state)} for the orbiter and lander simulations is shown in Figs. \ref{fig:orbiterMarsErrorRatio} and \ref{fig:landerMarsErrorRatio}, respectively. The theoretical ratios for several relevant periods (as taken from Fig. \ref{fig:rangeRangeRateCompre}) are also indicated. For the lander simulations, the theoretical ratio is close to the value expected from a periodic signal at the Mars and Earth orbital frequencies. For the orbiter simulations, however, the ratios are a factor 5 higher than these analytical values (Fig. \ref{fig:orbiterMarsErrorRatio}). The correlations between $\mathbf{x}_{M}(t_{0})$ and the other parameters are low for both data types. %with an average absolute value of about 0.05 in both cases and no values greater than 0.40. 
Since the analytical value for $\epsilon_{r_{M},s}/\epsilon_{r_{M},\dot{s}}$ is much more closely attained in the lander estimation (Fig. \ref{fig:landerMarsErrorRatio}), the discrepancy in the case of orbits indicates that the once-per-orbit signature of the spacecraft continues to have a significant effect on the estimation of Mars' initial state when obtained through the use of Eq. (\ref{eq:hybridStateTransitionMatrix}).

Rotational properties and Love numbers of Mars as estimated from orbiter data both have formal error ratios of around 1.0 (Fig. \ref{fig:orbiterParameterErrorRatio}). This is comparable to the orbiter initial-state estimation ratios (see Fig. \ref{fig:orbiterErrorRatio}), and indicates that the primary signature of these parameters is derived from a once-per-orbit signature. The range data perform slightly worse than the analytical result would indicate, due to stronger correlations between the parameters. 

The error ratio for the gravity field coefficients $C_{l,m}$ and $S_{l,m}$ is shown in Fig. \ref{fig:orbiterNumericalGravityFieldError}. These results show a clear trend in their formal error ratios, with $C_{l,l}$ and $S_{l,l}$ having a ratio about twice that of the initial state $\mathbf{r}_{0}$, increasing to roughly $(l+1)$ times that of the initial state for $C_{l,0}$. {Under the assumptions of the model outlined in Section \ref{sec:snr}, this would indicate that these parameters induce a periodic signature with a frequency of twice (for $m=l$) to $(l+1)$ times (for $m=0$) the orbital frequency.} 

{Since the orbits we have used are close to polar, the time-behaviour of the perturbation due to a given gravity field coefficient is mostly determined by $\boldsymbol{\nabla} P_{l,m}(\sin\phi)$, with $P_{l,m}$} associated Legendre polynomials at degree $l$ and order $m$, and $\phi$ the body-fixed latitude of the spacecraft \citep[\emph{e.g.} ][]{MontenbruckGill2000}. A spherical harmonic coefficient at this degree and order causes an acceleration on the spacecraft that is linearly proportional to $P_{l+1,m+1}$,  $P_{l+1,m}$ and $P_{l+1,m-1}$. For $P_{l,m}$, the number of zero crossings over a full orbit is 2 for $l=m$, while it is $l$ for $m=0$ and $m=1$. 

{Our results in Fig. \ref{fig:orbiterNumericalGravityFieldError} show that the influence of a function $P_{l,m}$ may be simplified to a sine function with the same number of zero crossings, for the purposes of our analytical comparison.} This allows the analytical criterion in  Fig. \ref{fig:rangeRangeRateCompre} to be used for approximating the relative contribution of range and Doppler data  to gravity field estimation. %A similar relation is observed for to the formal error ratios shown in Fig. \ref{fig:orbiterNumericalGravityFieldError}: ratio is approximately double that of the once-per-spacecraft orbit signature for $l=m$, and $l$ to $l+1$ times this once-per-orbit value for $l=m$. 
%ComparinThis indicates that the analytical relation shown in Fig. \ref{fig:rangeRangeRateCompre} gives a reasonable approximation for the relative contribution of range and Doppler data  to gravity field estimation.

The error ratios of the parameters estimated in the lander scenario are shown in Fig. \ref{fig:landerParameterErrorRatio}. They clearly fall into two categories: those for which the formal error ratio is close to that predicted for signatures at Mars' rotational period (dotted black line; $\epsilon_{q,s}/\epsilon_{q,\dot{s}}\approx 0.07$), and those for Mars' orbital period (full black line; $\epsilon_{q,s}/\epsilon_{q,\dot{s}}\approx 10^{-4}$). All of Mars' rotational properties fall into the former category, as expected. The parameters impacting Mars' orbit ($\beta$ and $J_{2\astrosun}$)  fall into the latter category, but their formal error ratio is about a factor two smaller than expected from our analytical approximation (assuming the signal to be at Mars' orbital period). This difference is due to the correlation with the Mars initial velocity parameters, which is $<0.2$ for both $\beta$ and $J_{2\astrosun}$ in the range-only simulations, and 0.6-0.7 for the Doppler-only simulations. %This indicates a clear limitation of the purely analytical approach: any significant change in correlations that are obtained with different data types will not be capture by this method.
Finally, one of the components of the lander position is estimated poorly using Doppler data only, an effect well known and discussed by \emph{e.g.} \cite{LeMaistreEtAl2013}. %The good determination of the lander position $x$-component, when using Doppler data, is due to a comparatively small correlation with $\beta$ in the range-only case (0.1), compared to the Doppler-only case (0.75).

%The results shown in this section are all for the case C simulation settings given in Table \ref{tab:typicalMissionsPositionVelocityCompare}. The results of the more limited case A and case B simulations are very similar to the results given in Figs. \ref{fig:orbiterErrorRatio}-\ref{fig:orbiterNumericalGravityFieldError}, and we do not discuss them explicitly.

\section{Discussion}
\label{sec:discussion}

%Based upon the results discussed in Section \ref{sec:results}, we can draw general quantitative conclusions on the comparative performance of an ILR system and existing radio Doppler systems for a range of scientific parameters of interest in planetary missions. %Despite the fact that our approach will only be valid as a first-order approximation (as discussed above), it provides a good initial estimate of the physical parameters and associated science goals for which the use of laser ranging can offer competitive performance. 

Having shown the results of our analytical and numerical analyses in Section \ref{sec:results}, we now discuss the potential application of ILR in planetary missions. %We start in Section \ref{sec:observationUncertaintyComparison} by discussing the manner in which to interpret the ratios $\Xi$ and $\sigma_{s}/\sigma_{\dot{s}}$ that we have obtained. Subsequently, we discuss the potential of ILR in spacecraft orbit determination and the determination of gravity fields in Section \ref{eq:gravityFieldScience}, for tidal and rotational characteristics in Section \ref{eq:rotationTidalScience} and ephemerides in Section \ref{sec:ephemeridesResults}. Finally, we discuss the potential for one-way laser ranging in Section \ref{sec:oneTwoWayScience}.

%\begin{figure*}[tb!]
%\centering
%\subfigure[]{
%\includegraphics[width=0.5\textwidth]{ParameterFormalErrorRatioRangeWrtDoppler_EstimationCase4OrbitCase_1.eps}
%}
%\caption{}
%\label{fig:orbiterNumericalParameterError}
%\subfigure[]{
%\includegraphics[width=0.45\textwidth]{LanderParameterFormalErrorRatioCase_0EstimationCase_2.eps}
%}
%\caption{.}
%\label{fig:landerNumericalParameterError}
%\end{figure*}

\subsection{Observation Uncertainty Comparison}
\label{sec:observationUncertaintyComparison}
Section \ref{sec:analyticalResults} gave the results of a conceptual criterion for comparing range and Doppler data using an analytical formulation. This was shown in Section \ref{eq:numericalResults} {to approximate the results of the numerical covariance analysis reasonably well}. %The analytical model starts to break down if the correlations change drastically between the range and Doppler simulations, with differences of a factor $>2$ observed for a few cases. The major discrepancy between numerical analytical results occurred for the estimation of Mars' initial state from orbiter data, where the analytical approach overestimated the relative strength of ILR by a factor up to 5. This is due to the strong signature of the orbiter's dynamics in this estimation. 
However, both the analytical method and the covariance analysis suffers from common limitations: they ignore any differences in the measurement uncertainty probability distributions, and dynamical model errors are neglected. 

%As discussed in Section \ref{sec:radiometricTrackingDataQuality}, 
Doppler tracking is close to being bias-free, and in the case of dual-frequency tracking, it has a noise spectrum that is close to Gaussian (Section \ref{sec:radiometricTrackingDataQuality}). For laser tracking, the single-shot uncertainty distribution will {be defined by the Gaussian pulse profile, convoluted with the detector impulse response} in the case of single-photon detection (Section \ref{sec:inherentStochastic}). {Both the pulse profile and the detector response can be characterized to high accuracy (Section \ref{eq:measErrors}).} However, as is the case in SLR/LLR, biases and instabilities at the several mm-level will likely continue to be an issue (Section \ref{sec:ilrTotalError}). %Depending on the temporal behaviour of these biases, their influence may be relatively small. %In exceptional cases, however, this temporal behaviour may correlate with the signature of estimated parameters \citep{DirkxEtAl2014b}, degrading the estimation.

The issue of stability will be especially significant for lander tracking, where model uncertainties are expected to be the dominant source of estimation error (Section \ref{sec:ilrTotalError}). For orbiter tracking, the dynamical model error will be more significant (compared to landers) and the impact of the precise observation noise spectrum less pronounced. 

\cite{DirkxEtAl2014} analyzed the impact of unresolved constant ILR biases for the Phobos Laser Ranging \citep[PLR; ][]{TuryshevEtAl2010} mission. They used 5 mm Gaussian measurement noise, and 5 mm constant unresolved bias, and obtained results indicating that the biases contribute about an order of magnitude more uncertainty to the estimated parameters than the Gaussian noise. For biases that vary quasi-randomly from pass to pass, this impact will in part average out. However, it will still cause the range-only simulations presented here to be more optimistic than the Doppler-only simulations, especially for the lander case. As such, any results from the criterion in Eq. (\ref{eq:approximateAnalyticalRatio}) that indicate similar signatures on Doppler and range data should be interpreted as indicating Doppler data will likely continue to be the better choice of data type. Nevertheless, reducing system and observation biases in laser ranging systems is a continuous and ongoing priority in ILRS activities.

%For a specific mission scenario, a full and detailed understanding can only be obtained through numerical analysis that includes a realistic error spectrum of the observables, and a representative difference between the truth and estimation models.

\subsection{Gravity Fields and Orbit Determination}
\label{eq:gravityFieldScience}

Radio-range measurements are poorly suited for estimating planetary gravity fields. The Doppler data are used as the primary input data type for current missions \citep[\emph{e.g.} ][]{MartyEtAl2009,KonoplivEtAl2011,MazaricoEtAl2014b}. %Estimating a gravity field up to degree and order $l$ entails the extraction of signatures in the tracking data at up to $(l+1)$-cycles per orbit. 
The results in Section \ref{sec:analyticalResults} indicate that the use of ILR becomes competitive for signatures with a period of 0.33-1.65 hours for range data precisions of 2-10 mm (corresponding to range data accuracy at approximately sub-mm to several-mm level, see Section \ref{sec:observationUncertaintyComparison}). The analytical analysis was shown  in Section \ref{eq:numericalResults} to provide a good approximation to the full numerical simulations for gravity field estimation (see Fig. \ref{fig:orbiterNumericalGravityFieldError}). This indicates that Doppler data are the superior choice for gravity field estimation, with the possible exception of very low gravity field degrees. For low degrees (2-4), ILR could meaningfully complement the Doppler data, which would require exceptional stability of the range data. {Since Doppler data are essentially bias-free, the competitive estimation of low-degree gravity fields will only be achievable by means of ILR if it is similarly (close to) bias-free (Section \ref{sec:observationUncertaintyComparison}).}

%For JUICE, for instance, the orbital period around Ganymede during the 500 km circular orbit phase will be about 3 hours \citep{GrassetEtAl2013}. The main BepiColombo spacecraft \citep{BenkhoffEtAl2010} will have a similar orbital period around Mercury (about 2.6 hours at a moderate eccentricity of about 0.16). 

%Assuming an orbital period of 3 hours, the 3 mm range measurement has similar sensitivity as Doppler case $A$ at degree 2, and the same or higher sensitivity up to degree 6 for Doppler case $B$. For the 6 mm range measurement, Doppler case $A$ is clearly superior even at degree 2, whereas Doppler case $B$ has the same sensitivity at degree 3. For higher degrees, the reduction in gravity field signal wavelength means that the Doppler measurement will yield superior results. However, in light of the presence of unresolved biases in ILR data at the mm-level (Section \ref{sec:observationUncertaintyComparison}), it is likely that ILR will not be a competitive method, even for low degrees. 

{The comparatively poor sensitivity} of ILR data to gravity field coefficients shows that a laser-only tracking system is unlikely to be a suitable design choice for a planetary orbiter. Any uncertainty in the target body gravity field will propagate into an increased error in the orbit determination of the spacecraft {\citep[\emph{e.g.} ][]{MazaricoEtAl2012}. %ILR will not be sensitive to any but the lowest order coefficients, so that 
Even if ILR is used} on an orbiter for other applications, it should always be in tandem with a Doppler tracking system, which can be used to measure the high-frequency variations in the spacecraft dynamics, chiefly the influence of the target body's gravity field. One possible exception is in the case of lunar missions. The gravity field of the Moon has been determined to such extreme accuracy \citep{LemoineEtAl2014} that the impact of its uncertainty on a typical spacecraft's orbit determination will likely be negligible. %Even in this case, uncertainties in non-conservative force modelling will limit the orbital accuracy that can be attained (Section \ref{sec:modelErrorSources}). %These errors are in part mitigated by estimation of correction coefficients/empirical accelerations.

%For spacecraft orbiting larger bodies the orbital period will be longer, increasing the potential value of laser ranging. The period of the science orbit of Juno around Jupiter will be approximately 10 days at an eccentricity of 0.95, for instance \citep{Matousek2007}. %When considering only the orbital period, it may be assumed that laser ranging could offer superior performance for signals up to moderate gravity field coefficients. 

\subsection{Rotational and Tidal Characteristics}
\label{eq:rotationTidalScience}

%Properties of the rotation and tidal deformation of solar system bodies may be retrieved directly from the motion of orbiters around these bodies \citep[\emph{e.g.}][]{MartyEtAl2009,MazaricoEtAl2014b}. 
For tidal parameters, the  signature on orbiter dynamics will show a combination of the spacecraft's orbital period, the rotation rate of the {body being orbited} and the period of the tidal forcing. The results in Fig. \ref{fig:orbiterParameterErrorRatio} show that it mainly is the once-per-spacecraft-orbit signature that is dominant in determining the formal error ratio $\epsilon_{s}/\epsilon_{\dot{s}}$. %The analytical criterion in Fig. \ref {fig:rangeRangeRateCompre} overestimates the relative strength of the range data in estimating tidal parameters by only 10-25 \%.

%\begin{figure*}[tb!]
%\centering
%\includegraphics[width=0.97\textwidth]{LanderStatePartials.eps}
%\caption{Normalized partial derivatives of Doppler (top) and range(bottom) w.r.t. Mars initial position components as a function of time. The long gaps in data are due to the solar occultations, where no data is simulated. Note that the normalizations are different for the range and Doppler data.}
%\label{fig:landerStatePartials}
%\end{figure*}

Similarly, rotational variations such as librations can have a broad range of periods, but significant variations typically do not have a period that is much smaller than the rotational period \citep[\textit{e.g.}, ][]{KonoplivEtAl2006,PetitEtAl2010}. %Variations with much larger periods, such as the precession rate, can be used to deduce the body's polar moment of inertia, which is used to constrain interior structure models \citep[\emph{e.g. }][]{KhanConnolly2008}. 
As with tidal parameters, our results in Fig. \ref{fig:orbiterParameterErrorRatio} indicate that it is again the once-per-spacecraft-orbit signature that is dominant in determining $\epsilon_{s}/\epsilon_{\dot{s}}$. For most tidal and rotational parameters, the analytical approximation ($\approx$0.95) slightly overestimates the strength of the range data, as Fig. \ref{fig:orbiterParameterErrorRatio} shows formal error ratios $\approx$ 0.9$-$1.25 (note that a higher ratio indicates a relatively weaker contribution of the range data).

%As discussed in the Section \ref{eq:gravityFieldScience}, the results in Section \ref{sec:results} show roughly similar strong signatures on laser and Doppler systems for spacecraft once-per-orbit signals. 

This indicates that orbiter ILR data could be used to estimate tidal and rotational characteristics at a level that is comparable to that from Doppler data. Doing so will require a low level of unresolved systematic error in the range data, at the several mm level, assuming that (non-conservative) force modelling on the spacecraft does not limit the estimation quality. Moreover, as discussed in Section \ref{eq:gravityFieldScience}, the extraction of signals from the dynamics of orbiters will require the inclusion of a Doppler system, to prevent the uncertainty in short-periodic perturbations from degrading the quality of the estimation results. As a result, ILR could be used to supplement Doppler tracking in the determination of rotational and tidal characteristics from orbiter dynamics, but would not be the optimal choice as a dedicated system.

Estimation of rotational properties from lander data is shown in Fig. \ref{fig:landerParameterErrorRatio}. Our analytical model closely predicts the $\epsilon_{s}/\epsilon_{\dot{s}}$ ratio, at the once-per-Martian-day frequency. Compared to orbiters, lander missions are more favorable for ILR, as the rotation period of a body (\emph{e.g.} its day) is typically longer than the orbital period of a spacecraft. %Similarly, tidal deformation of bodies will in most cases manifest itself most strongly as a once-per-revolution effect on the range to a lander, leading to a similarly strong signature on ILR data. 

Considering the typical rotational periods of bodies in our solar system, we can confidently state that laser range measurements to landers will be better suited for the estimation of rotational parameters than Doppler measurements will be. Fig. \ref{fig:landerParameterErrorRatio} shows a factor 20 improvement of Mars rotational parameters from ILR compared to Doppler data. For some fast-rotating bodies such as Phobos, this factor would be reduced to $<$10. For bodies with much slower rotational periods, such as Ganymede and Mercury, the formal error from ILR could be $>$100 and $>$1,000 times smaller than from Doppler data. However, at these levels of observational accuracy, many models will need to be improved to make full use of the data quality that would be available  {\citep{DirkxEtAl2014}}. %This makes it doubtful that the quality of the final accuracy of the science product from a laser ranging mission to Ganymede or Mercury would indeed be 70 and 1,000 better than that from a system using Doppler tracking. 

Nevertheless, our results clearly show the exceptional strength that {ILR} can have in characterizing rotational motion and tidal deformation in the solar system, especially for landers. Similarly, {it} will be well suited for the determination of the $h_{2}$ and $l_{2}$ Love numbers (and possibly $k_{2}$, due to its influence on rotational dynamics \citep[\emph{e.g.} ][]{WilliamsEtAl2001}).

\subsection{Solar System Ephemerides}
\label{sec:ephemeridesResults}
%For the determination of planetary ephemerides, radiometric range and VLBI measurements are currently the primary sources of spacecraft-based data (Section \ref{sec:preliminaryTypeComparison}). Laser range measurements will be about 30-60 times more accurate than range measurements from next-generation radio tracking systems ). %However, the use of laser ranging will not improve the angular position observations of spacecraft, which are obtained by VLBI measurements. 

%For VLBI observations, the state-of-the-art measurement accuracy of 1 nrad leads to an approximate position accuracy of 150 m for every AU of distance between Earth and the target. Although the VLBI measurements provide much looser constraints in the estimation of ephemerides than laser ranging will, its unique sensitivity to the out-of-plane component will continue to be an important asset for anchoring planetary ephemerides, especially over longer time scales.  %We discussed the potential of ILR measurements to improve planetary ephemerides in a more general sense in Section \ref{sec:discModelErrors}.

Fig. \ref{fig:marsErrorRatio} indicates that ILR is preferred over Doppler data for the determination of planetary ephemerides, which is unsurprising considering the current role of radio range data (Section \ref{sec:planetaryMissions}). The relative contribution of range and Doppler data to the estimation of ephemerides is well approximated by the criterion in Eq. (\ref{eq:approximateAnalyticalRatio}) for the case of lander missions. For orbiter data, Eq. (\ref{eq:approximateAnalyticalRatio}) overestimates the relative contribution of the range data by a factor of up to 5. Clearly, the Doppler data continues to contribute to the determination of planetary ephemerides, {by accurately extracting} the orbiter dynamics from the observations. As a consequence, ephemeris determination will benefit from the combination of Doppler and ILR data, for reasons discussed in Section \ref{eq:gravityFieldScience}. Fig. \ref{fig:landerParameterErrorRatio} shows that the parameters estimated jointly with ephemerides (here only $\beta$ and $J_{2,\astrosun}$) will benefit greatly from the use of ILR. These parameter are crucial in relativistic experiments \citep{Will2014}. For the case of landers, their uncertainty is reasonably approximated by the analytical formulation. {However,} properly decorrelating these parameters using ILR data {may} require laser data to multiple targets.

Due to the scarcity of ILR data when it will be first implemented, there will be an imbalance of several orders of magnitude between the measurements used for {creating solar system ephemerides}. For instance, when using ILR data from an Earth-Mars link, there will be mm-level range measurements for Earth and Mars, m-level (radiometric) range measurements for other solar system bodies at which orbiter/lander tracking data is available and km-level range measurements (radar) to bodies where no data from spacecraft tracking techniques is available. For many small bodies, no range data will be available at all, and ephemeris generation must be performed from astrometric data alone. %Additionally, dynamical model error, such as asteroid mass uncertainty, will degrade the ephemeris generation quality. 
{This effect, as well as dynamical model error (Section \ref{sec:modelErrorSources})} will degrade the fidelity of the orbit {estimation} of the bodies between which an ILR link is set up. Quantifying the exact requirements for a mission profile, tracking schedule, estimation settings, \emph{etc.} for optimally exploiting ILR data in ephemeris generation will require a dedicated study.

\section{Conclusion - The Science Case for ILR}
\label{sec:conclusions}
The purpose of the article has been twofold. Firstly, we have given a detailed overview of the sources of uncertainty in both the realization and analysis of ILR data (Section \ref{sec:lrData}). Secondly, we have compared the performance of ILR and radio Doppler data both analytically and numerically, to clarify the science case for ILR (Sections \ref{sec:compCrit}-\ref{sec:discussion}). 

Mm-precision normal points are feasible for ILR. %various issues in both the measurement and analysis process will prevent the data from being exploited to 
%However, the same level of accuracy will be challenging to attain 
Sub-cm accuracy of the data will be attainable, but reaching the mm-level accuracy is hindered by both measurement instabilities and model uncertainties, similar to SLR/LLR (Section \ref{sec:ilrTotalError}).  %Many small observation biases in SLR/LLR carry over directly to ILR, although the reflector signature is absent in transponder ranging, allowing the pulse temporal intensity-distribution at the receiver to be more accurately modelled. Therefore, operating at the single-photon level will mitigate part of the stability issues. 
%The impact of dynamical modelling errors on the science products is strongly dependent on the mission under study. For orbiters, non-conservative force modelling will be the main challenge, while for planetary ephemerides the uncertainties in masses and ephemerides of small bodies will limit the attainable model accuracy. %For the scientific interpretation of the products from ILR, other data sets and models may need to be improved as well. Relate the high quality estimated parameters (\emph{e.g.} tidal and rotational properties) to interior structure parameters will require additional improvements in data beyond

%We have exploited the relation between the range and Doppler observable to generate 
We have derived an analytical approximation of the sensitivity of ILR and radio Doppler data types, which is shown in Fig. \ref{fig:rangeRangeRateCompre}, under the assumption of sinusoidal signatures on the data. This figure indicates that the signatures with a period of 0.33-1.65 hours can be observed in the ILR and radio Doppler data at a similar signal-to-noise level. The results of our numerical covariance analysis largely validate the analytical approach, allowing Fig. \ref{fig:rangeRangeRateCompre} to be used as a conceptual design tool. %Appendix \ref{ref:semiAnalyticalResults} shows how the approach can be extended to eccentric orbits by a suitable change of variables.

However, instabilities in the range data accuracy, which are not directly included in the simulations, will limit the performance of ILR data to the upper bounds of the 0.33-1.65 hour range. That is, ILR will start to be competitive for determining signatures of $>$1.5-2 hours. For effects with a longer period than several hours, ILR data unambiguously provide a more accurate estimation, while Doppler data will continue to be the optimal data type for short-periodic effects. 
%Our analyses show that ILR could be used as a complementary data type in orbiter tracking, but that its 

ILR's weak sensitivity to short-periodic effects makes it a poor choice for gravity field estimation, {with the possible exception of low-degree coefficients}. Any orbiter with typical orbit determination requirements will continue to require Doppler data (with the possible exception of lunar missions). For typical mission profiles, ILR and Doppler data have a similar sensitivity to once-per-orbit effects (period of 1.5-2 hours). Therefore, ILR will be valuable in complementing the Doppler data for the estimation of tidal and rotational characteristics from orbiter tracking. As expected, ILR is exceptionally well suited to generating ephemerides, although our analytical approximation overestimates its strength by a factor 5 for the case of orbiter tracking.

In addition, when used for lander tracking, ILR will be an excellent method for the estimation of both tidal and rotational characteristics of the target body. For a Mars lander, ILR data produces estimates that are about a factor 10 more accurate than {the estimation from} Doppler data. The strength of ILR improves even more for bodies with slower rotation rates. %The distinction with orbiters is a direct consequence of the longer rotation period of Mars, compared to the spacecraft orbital period. 
%For lander scenarios, the formal error ratio of Doppler and ILR is predicted by Fig. \ref{fig:rangeRangeRateCompre} quite well. %For landers tracking data, the estimation quality of planetary ephemerides (and associated parameters such as $\beta$ and $J_{2,\astrosun}$) will be directly improved by ILR, as radio-range data is presently used as a key input. For these parameters, the formal error ratio fo ILR and Doppler is also accurately predicted by Fig. \ref{fig:rangeRangeRateCompre}. %For the long-period parameters, range data results in reduced correlations, further improving the science case for ILR on lander missions.

%The above conclusions are valid for two-way ILR data. One-way data suffers from clock noise, requiring the estimation of clock parameters during the estimation, which correlate strongly with long-periodic effects. A one-way system could complement radio tracking data, but providing significant improvements on radio data will be challenging. Doing so will require exceptional clock stability on the space segment.

We have focused on the estimation quality of orbits and geodetic parameters, concluding that the science case for landers is excellent, and it could serve a complementary role for orbiters. Owing to the highly accurate data that ILR will deliver, improvements in the various models entering not only the analysis, but also the interpretation of the estimation results, must be brought to a level where all data can be used to their full potential. Not only will this require significant theoretical effort, but it implies that more accurate knowledge must be obtained of various quantities that cannot be obtained from tracking data alone. Examples of synergistic data are magnetic field, heat flow, geological and seismic measurements, which will be important for the full characterization of a body's interior structure and composition. By combining these data from next-generation space missions with ILR, the full set of these measurements can be exploited to their full potential, allowing the study of planetary interiors to be brought to the next level.

\section*{Acknowledgements}
The authors are indebted to Luciano Iess for discussions on the strengths and weaknesses of planetary range and Doppler data, and to three anonymous reviewers, whose comments improved the clarity, conciseness and completeness of the manuscript. Part of this work was performed in the FP7 ESPaCE project, financially supported by the EC FP7 Grant Agreement 263466. %Sections of this work were adapted from Chapters 2, 3, 8 and 9 of the Ph. D. thesis by \cite{Dirkx2015}.

\appendix
\section{Semi-analytical Approach - Eccentric orbits}
\label{ref:semiAnalyticalResults}
A key exception for which the assumptions of Section  \ref{sec:snr} will not hold is when a spacecraft orbits with a substantial eccentricity, \emph{e.g.}, Juno ($e\approx 0.95$ w.r.t. Jupiter), Messenger ($e\approx 0.7$ w.r.t. Mercury), Mars Express ($e\approx 0.57$ w.r.t. Mars). {To extend the method of Section \ref{sec:snr} to non-spherical orbits, 
%Although non-spherical Kepler orbits can be described analytically, the closed-form partial derivative calculation becomes cumbersome. 
we} continue to use the criterion of Eq. (\ref{eq:observableSnr}), but compute the partial derivatives from the orbits directly using the method of \citep{Moyer2005}, instead of imposing them to behave sinusoidically. We limit ourselves to taking $q$ as the initial position $\mathbf{r}_{0}$ of the spacecraft w.r.t. the planet it is orbiting.

\begin{figure}[tb!]
\centering
\includegraphics[width=0.47\textwidth]{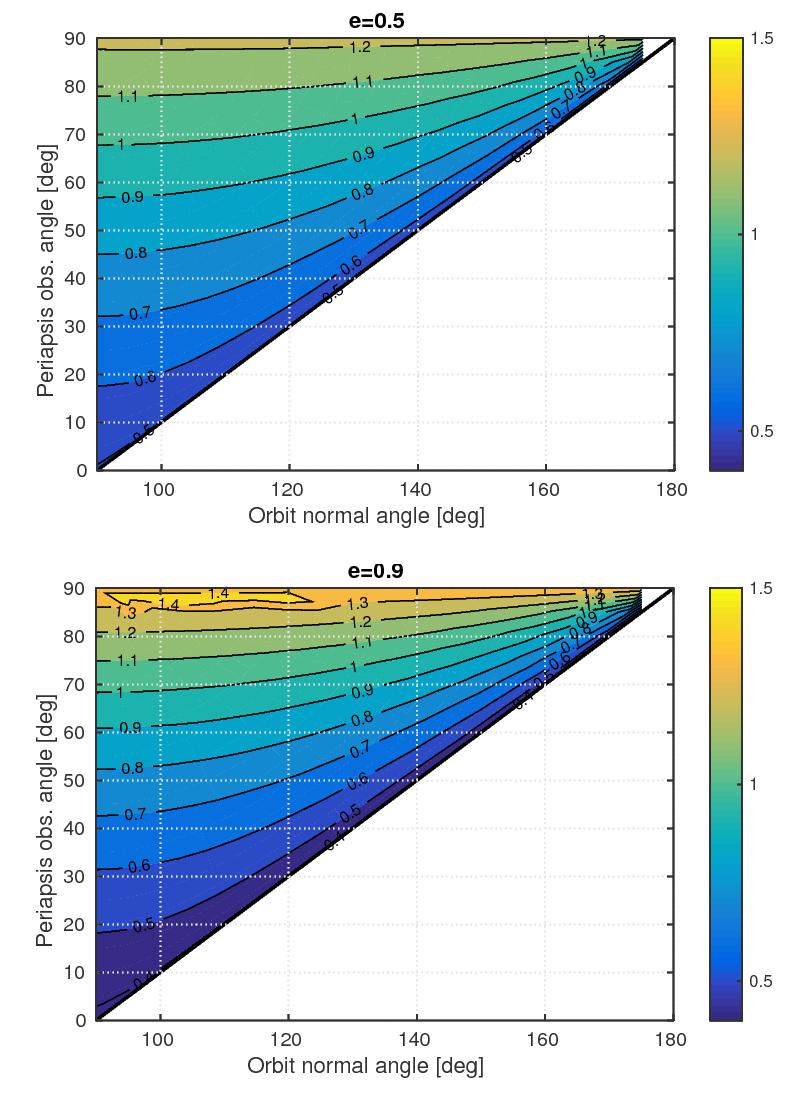}
\caption{Ratio of figures of merit $\Xi_{r_{0}}(e)/\Xi_{r_{0}}(e=0)$ for elliptical orbits and circular orbits as a function of viewing geometry. {The color scale denotes this ratio}. Note that the periapsis distance $r_{p}$ is set equal in the $e=0$ and $e\neq 0$ simulations.}
\label{fig:eccentricOrbitRelativeValue}
\end{figure}

We vary the eccentricity from 0 to 1 and assess the influence on the behaviour of %the partial derivatives and 
$\Xi_{q}$. %For a spherical orbit, where $e=0$ exactly, the signature on the data is independent of the true anomaly. 
For an eccentric orbit, %however, this is no longer the case and 
the influence of the geometry of the orbital plane w.r.t. the observation line-of-sight needs to be analyzed. We parameterize this geometric dependency by two angles: the angle between the line-of-sight vector and the spacecraft orbital plane, and the angle between the line-of-sight vector and the spacecraft orbital velocity vector at periapsis. %We do not include the semi-major axis and/or orbital period as additional parameters in our model, as our resuls will be presented as the ratio of Eq. (\ref{eq:rangeRangeRateCompareCriterion}) evaluated for a spherical orbit and that for the elliptical case.

%These of the relative influence of the data types shown in Fig. \ref{fig:rangeRangeRateCompre} only hold for circular orbits, omitting the influence of the eccentricity of these missions. Many planetary missions have substantial orbital eccentricity (Juno, Venus Express, Mars Express, BepiColombo, MESSENGER). 

Using the methodology outlined above, we have generated values of $\Xi_{q}$ (with $q$ the initial position $\mathbf{r}_{0}$, and computing the associated $\Xi_{r_{0}}$ as the root sum square of the constituent vector components of $\partial h/\partial\mathbf{q}$) for elliptical Kepler orbits under a full range of observational geometries w.r.t. the observer line-of-sight.

As a test case, we use a Mars orbiter for our analysis. When generating the values of $\Xi_{r_{0}}$ for two cases, one with $e=0$ and one with $e>0$, we find that using the same spacecraft semi-major axis for both does not lead to insightful results. Instead, when using the same spacecraft periapsis distance $r_{p}=a(1-e)$ for the two cases, the results for zero and non-zero eccentricities can be related much more intuitively. We show the results of this analysis in Fig. \ref{fig:eccentricOrbitRelativeValue}, were we plot the ratios of $\Xi_{r_{0}}(e)/\Xi_{r_{0}}(e=0)$, using the same $r_{p}$ to compute the two values of $\Xi_{r_{0}}$.

From this figure, it can be seen that even for substantial eccentricities, the ratio $\Xi_{r_{0}}(e)/\Xi_{r_{0}}(e=0)$ remains close to 1 for a broad range of eccentricities and observational geometries. In our simulations ($e\le$ 0.9) , the results deviate from the analytical ratio by less than 50 \% in all cases and less than 10 \% in most cases. This indicates that the analytical criterion shown in Fig. \ref{fig:rangeRangeRateCompre} continues to be largely applicable even for large eccentricities. %The results given in Fig. \ref{fig:eccentricOrbitRelativeValue} can be used to assess the validity of the analytical criterion for a given eccentric orbit. 
This result greatly simplifies the first-order analysis of the added value of an ILR system, as it allows the analytical results for spherical orbits to be used for arbitrary eccentricities.

%Fig. \ref{fig:eccentricOrbitRelativeValue} compares results for equal $r_{p}$, not equal orbital period ${\omega_{P}}$. Therefore, 
When using Fig. \ref{fig:rangeRangeRateCompre} to analyze the comparative strength of range and Doppler data for elliptical orbits with period $\omega_{P}$, the following scaling should be applied to obtain the value of $\omega$ at which Fig. \ref{fig:rangeRangeRateCompre}  is to be read:
\begin{align}
\omega=(1-e)^{-1.5}\omega_{P}\label{eq:adjustedPeriod}
\end{align}
so that for elliptical orbits the Doppler data continues to be the preferred method over ILR for a larger range of values of $\omega_{P}$.

\bibliographystyle{apalike} 
\bibliography{Bibliography}

\end{document}